A SINGLE-PROCESSOR APPROACH TO SPEECH PROCESSING PIPELINE OF

BILATERAL COCHLEAR IMPLANTS

by

Taher Shahbazi Mirzahasanloo

APPROVED BY SUPERVISORY COMMITTEE:

_________________________________________

Dr. Nasser Kehtarnavaz, Chair

_________________________________________

Dr. Carlos Busso

_________________________________________

Dr. John H. L. Hansen

_________________________________________

Dr. Issa M. S. Panahi



*To my grandmother*

*To my parents*

*To my sisters, Fatemeh and Zahra*

*and*

*To my brothers, Nasser and Mansour*

A SINGLE-PROCESSOR APPROACH TO SPEECH PROCESSING PIPELINE OF

BILATERAL COCHLEAR IMPLANTS

by

TAHER SHAHBAZI MIRZAHASANLOO

DISSERTATION

Presented to the Faculty of

The University of Texas at Dallas

in Partial Fulfillment

of the Requirements

for the Degree of

DOCTOR OF PHILOSOPHY IN

ELECTRICAL ENGINEERING

THE UNIVERSITY OF TEXAS AT DALLAS

August 2014

ACKNOWLEDGMENTS


All the years of my PhD studies at The University of Texas at Dallas would not have been such a wonderful and enjoyable experience if it were not for all my amazing friends, colleagues, advisers and family that I have been very fortunate to always have around. I would not have been able to accomplish this dissertation work without the help and support that I received from them. Acknowledging their help is a great pleasure for me and will always remind me of their contributions to this project. I am truly grateful.

Dr. Vanishree Gopalakrishna helped me get started on the PDA project and always patiently provided her feedback whenever I needed. Many thanks go to Hussnain Ali, Milad Omidi and Dr. Nima Yousefian from the Cochlear Implant Laboratory for their technical helps. Hussnain helped me a lot on programming the PDA and was always responsible for the software issues we would often face. It is nice to have colleagues of his knowledge, skills and responsibility in the team. Nima guided me to setup my experiments with dual-microphone data and design the bilateral cochlear implant experiments. Not only a knowledgeable colleague and a helpful critic on the technical aspects of the project, he has also been a great source of help during my defense and preparation of the dissertation, as well as a real assistance after leaving UTD. I owe him a lot. I am thankful to Milad whose comments and critics during the first stages of my research, made a great impact. He is a skillful engineer and researcher. I would also like to thank Masoud Farshbaf, from the Quality of Life Technology Laboratory, who was always the first person for me to ask for help at the tough times of debugging the PDA code. A bunch of thanks go to




Mohammad, Ali, Rasoul, Masoud and all my friends who helped me a lot in my proposal and final exams. I would like to thank my fellow Signal and Image Processing Laboratory members who created a friendly environment in the lab and made working there, enjoyable and fun: Dr. BoRam Kim, Dr. Siamak Yousefi, Sidharth Mahotra, Chandrasekhar Patlolla, Sang Jae Nam, Kui Liu, Chih-Hsiang Chang, Chen Chen, Reza Pourreza and Kimia Saki. I have been enjoying working with Kimia on the continuation of this research; she is a very motivated and hardworking researcher. I would also like to give thanks to Dr. Soroosh Mariooryad from the Multimodal Signal Processing Laboratory who was a big help in the preparation of my dissertation despite coinciding with his busy time of working on his own dissertation.

Special thanks are due to the committee members of the dissertation, Dr. Herve Abdi, Dr. Carlos Busso, Dr. John Hansen, and Dr. Issa Panahi for the very interesting discussions and comments in the proposal presentation and the final exam, as well as for their invaluable feedbacks after the defense which significantly improved the dissertation. It was a great honor to have them in my dissertation committee. I would also like to thank the Electrical Engineering department staff, especially Ms. Rosarita Khadij Lubag, Ms. Suzanne Newsom and Ms. Mary Gribble for their assistance with my several technical presentations in the department, research assistantship paperwork and my academic coursework registrations. A better presentation of the dissertation was obtained after formatting corrections suggested by Ms. Amanda Aiuvalasit from the Office of Graduate Studies. I also express my appreciation for her comments to make the document read better.

It was one of the most painful moments in my life to learn that UTD lost the departed professor, Dr. Philip Loizou whose impact on the cochlear implant research community and his influence



on the lives of thousands of patients cannot be overestimated. I cannot emphasize more how wonderful it was to have the honor of being his student. He will long be remembered for his impact on all of us.

The accomplishment of this work would have undoubtedly been impossible without the continuous support that I received from my advisor, Dr. Nasser Kehtarnavaz. His passion about his research fields, his extensive knowledge and his generous devotion of his time and energy to guide me in my research challenges, provided me with strong motivations. I am deeply grateful for everything. I would also like to express my gratitude one more time to Dr. Babak Nadjar Araabi, and Dr. Majid Nili Ahmadabadi from University of Tehran for their invaluable support during my transition to UTD. I consider myself very fortunate to have the honor of having had them as my supervisors.

Definitely much more than what I have done, completion of my PhD and this dissertation work are largely due to my family. Their kind, generous and indescribable support of my studies while having me away from home during these years, cannot be thanked enough. Nothing more can be mentioned as I cannot express how much I love them.

This work was partially supported by the grant no. DC010494 from National Institutes of Health (NIH).

July 2014



PREFACE

This dissertation was produced in accordance with guidelines which permit the inclusion as part of the dissertation the text of an original paper or papers submitted for publication. The dissertation must still conform to all other requirements explained in the "Guide for the Preparation of Master's Theses and Doctoral Dissertations at The University of Texas at Dallas." It must include a comprehensive abstract, a full introduction and literature review, and a final overall conclusion. Additional material (procedural and design data as well as descriptions of equipment) must be provided in sufficient detail to allow a clear and precise judgment to be made of the importance and originality of the research reported.

It is acceptable for this dissertation to include as chapters authentic copies of papers already published, provided these meet type size, margin, and legibility requirements. In such cases, connecting texts which provide logical bridges between different manuscripts are mandatory. Where the student is not the sole author of a manuscript, the student is required to make an explicit statement in the introductory material to that manuscript describing the student's contribution to the work and acknowledging the contribution of the other author(s). The signatures of the Supervising Committee which precede all other material in the dissertation (or thesis) attest to the accuracy of this statement.



A SINGLE-PROCESSOR APPROACH TO SPEECH PROCESSING PIPELINE OF

BILATERAL COCHLEAR IMPLANTS

Publication No. _____________________

Taher Shahbazi Mirzahasanloo, PhD
The University of Texas at Dallas, 2014

Supervising Professor:  Dr. Nasser Kehtarnavaz


This dissertation covers a single-processor approach to the speech processing pipeline of bilateral Cochlear Implants (CIs). The use of only a single processor to provide binaural stimulation signals overcomes the synchronization problem, which is an existing challenging problem in the deployment of bilateral CI devices. The developed single-processor speech processing pipeline provides CI users with a sense of directionality. Its non-synchronization feature as well as low computational and memory requirements make it a suitable solution for actual deployment. A speech enhancement framework is developed that incorporates different non-Euclidean speech distortion criteria and different noise environments. This framework not only allows the design of environment-optimized parameters but also enables a user-specific solution where the anthropometric measurements of an individual user are incorporated into the training process to obtain individualized bilateral parameters. The developed techniques are primarily meant for bilateral CIs, however, they are general purpose in the sense that they are also applicable to




binaural hearing aids, bimodal devices having hearing aid in one ear and cochlear implant in the other  ear as well as dual-channel speech enhancement applications. Extensive experiments have shown the effectiveness of the developed solution in six commonly encountered noise environments compared to a similar one-channel pipeline when using two separate processors or when using independent sequential processing.



TABLE OF CONTENTS













LIST OF FIGURES









LIST OF TABLES









INTRODUCTION

Cochlear Implants (CIs) are prosthetic devices that are used to restore hearing sensation in profoundly deaf people. Speech understanding by CI users has been reported to be acceptable in quiet and controlled listening conditions, but in actual noisy environments, it has been shown to decrease significantly (Remus and Collins 2005; Fetterman and Domico 2002). This issue has led to the development of speech enhancement algorithms to suppress noise such as the ones in (Loizou 2006; Hu, et al. 2007; Loizou, Lobo and Hu 2005). These algorithms have involved noise reduction strategies either as a preprocessing step before delivering speech to the CI speech processing pipeline or devising a noise attenuation technique in a built-in function as a CI speech processing component (Loizou 2006).

Real-life experiences of CI users often include dealing with speech in noisy environments having different noise characteristics. An environment-adaptive speech enhancement capability would allow CIs to operate more effectively in dealing with different types of noisy environments. A system capable of adapting its parameters to different noisy environments was developed previously in (Gopalakrishna, et al. 2012). This system involved an environment-adaptive noise suppression technique which automatically selected a set of parameters trained offline for each noisy environment. As a result, automatic noise suppression was achieved by switching among appropriate parameters via a noise classification decision in an online manner. Also, the system design was done in such a way that some of its components shared the same computations, leading to a computationally efficient pipeline.





In unilateral CIs, there is no directional information perceived by users, causing difficulties for them to locate sound sources (Litovsky, et al. 2004; Ching, Van Wanrooy and Dillon 2007). Bilateral CIs provide a natural way to create a sense of directionality. There are many studies, e.g. (Kühn-Inacker, et al. 2004; Litovsky, Johnstone and Godar 2006; Litovsky, et al. 2004), and (Van Hoesel and Tyler 2003; Müller, Schon and Helms 2002; Van Hoesel 2004), supporting that wearing two CI devices, instead of only one, improves speech understanding.

Bilateral CIs, utilizing multi-microphone or multi-channel techniques, have been developed mainly based on adaptive beamforming algorithms (Kokkinakis and Loizou 2010). Limited attempts have been made in the literature towards developing strategies that are computationally efficient on resource-limited processors. This dissertation has been an attempt to advance the CI technology toward the actual deployment of a computationally practical speech processing pipeline in bilateral CI devices. Currently, there exist both hardware and software challenges for delivering synchronized binaural stimulation signals to the stimulation electrodes of bilateral CIs. A lack of synchronization causes the loss or distortion of localization cues, thus producing limited benefits to bilateral CI users. The use of a single processor naturally overcomes the synchronization problem (Mirzahasanloo, et al. 2013). However, the use of a single processor creates computational and memory challenges in the implementation of the bilateral CI speech processing pipeline.

In this dissertation, the use of a single processor feeding both the left and right CIs are considered to achieve an environment-adaptive speech enhancement pipeline for bilateral CIs. Using two processors would require the left and right signals to be synchronized. In the



developed approach, since the bilateral stimulation signals are provided by only one processor, the synchronization issue does not need to be addressed. In addition, the use of a single processor makes the entire system more cost-effective.

Previous works (Gopalakrishna, et al. 2010; Mirzahasanloo, et al. 2013; Gopalakrishna, et al. 2012; Erkelens, Jensen and Heusdens 2007) have considered the environment-adaptability aspect by optimizing different gain tables for different noise environments, but have not studied the effects of different optimization criteria for different noise types. The problem of bilateral speech enhancement using a single processor becomes more challenging when considering non-Euclidean distortion measures. A generalized optimization framework is thus introduced in this dissertation that allows the utilization of other distortion measures (Loizou 2005; Fingscheidt, Suhadi and Stan 2008). Specifically, the optimization is done for three most commonly used distortion measures consisting of the traditional Weighted-Euclidean (WE), Log-Euclidean (LE) and Weighted-Cosh (WC) (Erkelens, Jensen and Heusdens 2007; Loizou 2005; Fingscheidt, Suhadi and Stan 2008; Erkelens and Heusdens 2008). The solutions provided are general purpose in the sense that they incorporate different weights over reference and non-reference signals as well as different parameter weights. As a result, the solutions for the data-driven unilateral enhancement gain optimization based on the WE, LE and WC criteria become special cases of this generalized framework. Although three most commonly used distortion measures are considered in this dissertation, the discussed framework can be applied to any differentiable measure.



Environment-adaptability demands statistics information of actual noise data in different environments. For Minimum Mean Squared Error (MMSE) and log-MMSE based solutions (Ephraim and Malah 1984; Ephraim and Malah 1985), a specific noise distribution model is presumed in advance which may not be necessarily an effective model in a wide range of real noise environments. Also, the model has to be regarded as fixed in all the environments (Gopalakrishna, et al. 2012). Such solutions, though optimal, are prone to modeling inaccuracies and estimation errors of Signal to Noise Ratio (SNR) and noise statistics (Erkelens, Jensen and Heusdens 2007; Erkelens and Heusdens 2008). The developed framework is able to provide near-optimal solutions by relaxing modeling assumptions and estimating errors using data-driven techniques (Mirzahasanloo, et al. 2013). Basically, this framework enables designing enhancement models that outperform model-based solutions in real noise environments (Gopalakrishna, et al. 2012; Mirzahasanloo, et al. 2013). Also, it allows the utilization of different distortion criteria (Loizou 2007; Loizou 2005; Fingscheidt, Suhadi and Stan 2008) by using non-linear optimization techniques (Mirzahasanloo and Kehtarnavaz 2013a) for which finding the model-based optimal analytical solutions are quite challenging.

The framework in (Mirzahasanloo, et al. 2013) is generalized by using the approach in (Erkelens, Jensen and Heusdens 2007), where only the Euclidean criterion for the bilateral case was used without providing adequate spectral resolution for the Head-Related Transfer Function (HRTF) estimation. This extension leads to a transformation model that eliminates all time-domain delay estimations and exploits perceptual frequency groupings for memory-efficient solutions. Any other gain solution can also be easily integrated into this unified speech



enhancement system. For example, MMSE or log-MMSE optimal suppression gain (Loizou 2007; Ephraim and Malah 1984; Ephraim and Malah 1985) along with any other HRTF estimation filters can be used.

The developed framework can be scaled for any directional hearing enhancement applications in addition to bilateral cochlear implants (Mirzahasanloo, et al. 2013). Interaural Time Difference (ITD) binaural cues are usually lost in bilateral CIs due to the difficulties associated with achieving synchronized signals. This issue drastically limits the localization benefits that CI users could get by receiving bilateral implants. The capability of retaining such cues in the developed framework is expected to provide promising bilateral benefits in future CI devices. Similar issues exist in binaural hearing devices; cochlear implants with contra directional microphones to provide a sense of directionality and bimodal devices that use hearing aid in one ear and cochlear implant in the other. The generality of the solutions achieved in this dissertation are also applicable to such devices.

Along with speech enhancement developments for bilateral CIs, the reliability of the system is enhanced by taking advantage of the two (left and right) signal sources, leading to a more robust operation in noisy environments. The overall performance of the solutions depends not only on the noise suppression component, but also on the effectiveness of noise classification. An improvement of the noise classification component that was previously developed in (Gopalakrishna, et al. 2012; Mirzahasanloo, et al. 2013; Mirzahasanloo, et al. 2012) is also made. This improvement is achieved by using the signals from a dual-microphone instead of a single microphone. The solution reached maintains the computational efficiency aspect of the



previously developed pipelines. In addition, to quantitatively evaluate the overall performance of an entire pipeline, a new measure is defined (Mirzahasanloo and Kehtarnavaz 2013b).

In general, existing speech enhancement algorithms introduce different types of distortions (Loizou, 2007). These distortions often cause decreased intelligibility scores. Intelligibility of speech is of a major concern to CI users. Although improving quality of speech is important in high noise levels, it is often desired to leave speech unprocessed in lower noise levels in order to avoid processing distortions or maintaining speech intelligibility. A quiet detection capability is thus added to the previously developed cochlear implant pipeline in order to turn off the suppression component for very low-level noise or practically quiet energy frames.

Furthermore, the presence of music demands a different suppression approach compared to noisy speech. A mechanism is also added to distinguish music frames from noise frames so that if desired a different suppression approach is applied (Mirzahasanloo, Kehtarnavaz and Panahi 2013).

Chapter 1 provides an introduction to basic noise suppression concepts and speech enhancement algorithms along with a brief review of the unilateral environment-adaptive speech enhancement pipeline (Gopalakrishna, et al. 2012). Section 1.1 provides an introductory description of the most commonly used noise suppression algorithms within the context of speech enhancement. More details and different algorithms can be found in (Loizou 2007).

Section 1.2 describes the unilateral enhancement pipeline which is the pipeline used for the extension to bilateral CIs.



In this dissertation, the techniques of the automatic enhancement framework in (Gopalakrishna, et al. 2012) are extended to bilateral cochlear implants. The initial extension is presented in Chapter 2. The main challenges addressed are how to keep speech distortions as low as possible while at the same time not allow the computational complexity to increase.

Generalizations to non-Euclidean distortion criteria and the utilization of non-linear optimization techniques to solve the single-processor data-driven dual-channel noise suppression for use in the bilateral CI pipeline are discussed in Chapter 3. The three commonly used distortion criteria are studied in Section 3.1.

A unified dual-channel speech enhancement framework is presented in Chapter 4, which is highly customizable in terms of different noise environments, different speech distortion criteria, different specific suppression and HRTF gain parameters, where anthropometric measurements of an individual user can be used in the training process to obtain individualized bilateral parameters.

Chapter 5 presents the environment detection improvements for unilateral and bilateral CI pipelines including the dual-microphone noise classification improvement in Section 5.2, and adding quiet and music detection capabilities in Section 5.3.

The experimental results and their discussion are then included in Chapter 6.

# CHAPTER 1

# NOISE SUPPRESSION AND ENVIRONMENT DETECTION[*]

This Chapter presents an overview of statistical data-driven approaches to speech enhancement for CIs. In addition, the components of the environment-adaptive pipeline are mentioned.

## 1.1 Data-driven noise tracking and speech enhancement

Statistical spectral enhancement methods can be characterized based on three main items that follow (Loizou 2007).

### 1.1.1 Prior SNR estimation

Decision-directed approach is the most commonly used method (Ephraim and Malah 1984). Also, the modifications proposed in (Erkelens, Jensen and Heusdens 2007) address the bias aspect in the convergence behavior of the decision-directed rule.

Let $R_k(n)$ and $A_k(n)$ be the noisy and clean spectral amplitudes in the frequency bin $k$ for the time frame $n$, respectively. The clean amplitude estimates $\hat{A}_k$ are then derived by applying a Signal to Noise Ratio (SNR)-dependent gain function $G$ to the noisy amplitudes as follows:

$$\hat{A}_k = G(\zeta_k, \xi_k)R_k \tag{1.1}$$

---

[*]©(2013), ELSEVIER. Portions reprinted with permission from (Mirzahasanloo, T., N. Kehtarnavaz, V. Gopalakrishna, and P. Loizou. "Environment-adaptive speech enhancement for bilateral cochlear implants using a single processor." Speech Commun, 55, 2013: 523-534)





where $\zeta_k$ and $\xi_k$ denote the prior and posterior SNRs which are defined as $\zeta_k = \lambda_x(k) \big/ \lambda_d(k)$

and $\xi_k = R_k^2 \big/ \lambda_d(k)$ . The computation of these SNRs require estimates of the clean spectral

variance $\lambda_x(k)$ and the noise spectral variance $\lambda_d(k)$ . The decision-directed estimator uses the

following rule to update the prior SNR for each time frame $n$

$$
\begin{aligned}
\hat{\xi}_k(n) &= \frac{R_k^2(n)}{\lambda_d(k,n)} \\
\hat{\zeta}_k(n) &= \alpha \frac{\hat{A}_k^2(n-1)}{\lambda_d(k,n)} + (1-\alpha)\max(\hat{\xi}_k(n)-1, \zeta_{\min})
\end{aligned}
\tag{1.2}
$$

where $\alpha$ is a weight close to one (for the results reported later in Chapter 6, $\alpha$ was set to 0.98),

and $\zeta_{\min}$ is a lower bound on the estimated value of $\hat{\zeta}_k$ (for the results reported later in Chapter

6, $\zeta_{\min}$ was set to -19dB).

## 1.1.2   Reconstruction

Based on the estimated prior and posterior SNRs, the spectral amplitude of the enhanced signal is

retrieved from the noisy input signal using an assumed probability density function and the

optimization of an objective function. The optimization solution for the reconstruction gain

defined in (1.1) is derived either analytically or obtained using data-driven training algorithms.

Two noteworthy solutions include Minimum Mean Square Error (MMSE) and log MMSE

spectral amplitude estimations.



### *MMSE spectral amplitude estimation*

MMSE spectral amplitude estimation (Ephraim and Malah 1984) provides an analytical solution

for Gaussian density and optimizing the minimum mean-squared error, that is

$$G_{MMSE}(\zeta_k, \xi_k) = \frac{\sqrt{\pi v_k}}{2\xi_k} \Phi(-1/2; 1; -v_k) \tag{1.3}$$

where

$$v_k = \frac{\zeta_k}{1 + \zeta_k} \xi_k \tag{1.4}$$

and $\Phi(a; b; x)$ denotes the confluent hypergeometric function (Abramowitz and Stegun 1965).

### *MMSE log spectral amplitude estimation*

MMSE log spectral amplitude estimation (Ephraim and Malah 1985) provides an analytical

solution for Gaussian density and optimizing the minimum mean-squared log spectral error, that

is

$$G_{\log-MMSE}(\zeta_k, \xi_k) = \frac{\zeta_k}{1 + \zeta_k} \exp\{\frac{1}{2} \int_{v_k}^{\infty} \frac{e^{-t}}{t} dt\} \tag{1.5}$$

where $v_k$ is the same as in (1.4).

### *Maximum a posteriori amplitude estimation*

Maximum a posteriori (MAP) amplitude estimation approaches include joint MAP estimator

proposed in (Lotter and Vary 2005).

### *Data-driven approaches*

The data-driven approach utilized here is the one reported in (Erkelens, Jensen and Heusdens

2007). It uses a gain table representation over prior and posterior SNRs for the amplitude

estimation. The optimal gains corresponding to this lookup table representation are then obtained

by using a distortion measure defined over noisy and clean spectral data.



### 1.1.3   Noise estimation

In parallel to the prior SNR estimation and the amplitude reconstruction, noise statistics need to be estimated. Depending on whether the noise is stationary or non-stationary, appropriate rules can be adopted.

*Stationary noise*

For stationary noise, constant variance is usually estimated in the first few silent frames of a sentence.

*Non-stationary noise*

The minimum statistics method (Martin 2001) is often used for the non-stationary case (Erkelens and Heusdens 2008).

To design the enhancement system optimized for different environments, either the noise estimation or the reconstruction needs to be parameterized based on the noise data. As will be discussed in Section 2.1, the gain table representation in (1.1) is utilized for training the reconstruction function in this dissertation.

The approach in (Gopalakrishna, et al. 2012) as outlined in Figure 1.1 incorporates a gain table representation for the noise tracking transformation and uses the log-MMSE reconstruction in (1.5) for the final amplitude estimation along with the environment-specific noise estimation trained and optimized for various noise environments. The main idea is relying on the data-driven gain table for the noise tracking (Erkelens and Heusdens 2008). The reconstruction part is independent of the environment and the same algorithm is used for all the noise types.



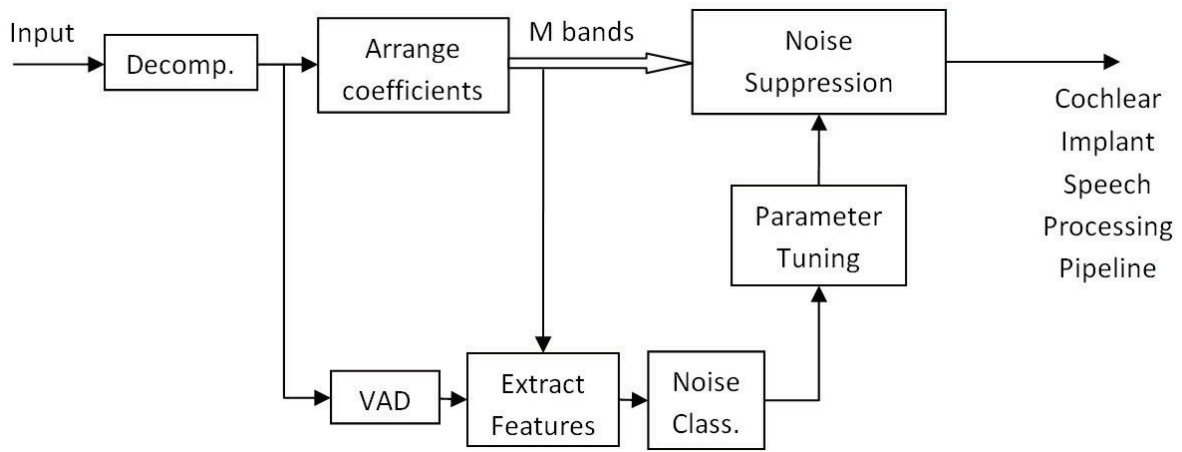

Figure 1.1. Block diagram representation of the previously developed unilateral cochlear implant pipeline.

## 1.2 Environment-adaptive noise suppression for unilateral cochlear implants

Figure 1.1 illustrates a block diagram of the environment-adaptive pipeline for unilateral cochlear implant speech processing that was previously developed in (Gopalakrishna, et al. 2012). This pipeline involves two main parallel paths for noise detection and noise suppression. A Voice Activity Detector (VAD) is used to determine signal frames containing speech. When it is purely noise, a Gaussian Mixture Model (GMM) classifier, trained based on a number of noise classes, is used to determine the noise type. After noise detection, the corresponding optimized gain parameters are loaded to a noise suppression function making the suppression path specific to the detected noise environment.

### 1.2.1 Environment detection

To characterize the noise frames for classification, a 26-dimensional feature vector which includes a combination of MFCCs (mel-frequency cepstral coefficients) with their first



derivatives is used providing high classification rates while not being computationally intensive. A total of 40 overlapping triangular filters are used to map the 64-frequency bands magnitude spectrum of the WPT (Wavelet Packet Transform) signal decomposition into 40 bins in mel scale. The first 13 filters are spaced linearly while the remaining 27 filters are logarithmically mapped by a discrete cosine transform generating 13 MFCCs.

In a previous study (Gopalakrishna, et al. 2010), an SVM (Support Vector Machine) classifier with radial basis kernel was used. However, to perform multiclass noise classification, the computational efficiency aspect of SVM poses a limitation and thus a GMM (Gaussian Mixture Model) with two Gaussians was utilized to provide a balance between classification performance and computational complexity. The parameters of the GMM classifier are estimated using k-means clustering and the Expectation-Maximization (EM) algorithm.

### 1.2.2   Noise suppression

In spectral domain, a gain function is assumed to be applied on magnitude spectrum of the input noisy speech signal providing an estimate of the associated clean spectrum. This gain is represented as a function of prior and posterior SNRs minimizing a mean squared error over a training set of noisy and clean sample pairs (Ephraim and Malah, 1985). This decision-directed approach is the most commonly used method to estimate the prior SNR. However, as discussed in (Erkelens, Jensen and Heusdens 2007), this approach leads to biased and erroneous results for some SNR values causing underestimation or overestimation of suppression. It is also worth mentioning that the gain function solution obtained using the MMSE and logMMSE estimators in (Ephraim and Malah 1985) assume specific distributions for the noise and speech spectra which may not necessarily be the best fitting distributions. To account for these modeling and



estimation shortcomings, a data-driven approach as proposed in (Erkelens, Jensen and Heusdens 2007; Erkelens and Heusdens 2008) is adopted where the gain values are obtained via a minimization formulation (Erkelens, Jensen and Heusdens 2007; Loizou 2005; Fingscheidt, Suhadi and Stan 2008). For non-stationary noise tracking, the tabular representation is considered to provide an estimation of noise spectrum. Then, this estimate is used in any analytic gain suppression function to provide the enhanced magnitude spectrum, e.g. the log MMSE estimator as used in (Erkelens, Jensen and Heusdens 2007).

The data-driven nature of the approach in (Gopalakrishna, et al. 2012) allows one to optimize the gain representation independently for each environment by considering the corresponding dataset. Because such a solution is MMSE optimal, it outperforms the conventional model-based methods (Erkelens, Jensen and Heusdens 2007). On the other hand, as different gain table parameters are optimized and used for different noise types, the overall performance becomes superior over that of a fixed noise suppression approach (Gopalakrishna, et al. 2012).

# CHAPTER 2

## SINGLE-PROCESSOR BILATERAL SPEECH PROCESSING PIPELINE[†]

In this Chapter, a speech processing pipeline in noisy environments based on a single-processor implementation is developed for utilization in bilateral cochlear implants. A two-channel joint objective function is defined and a closed-form solution is obtained based on the weighted-Euclidean distortion measure. This solution allows one to obtain environment-optimized parameters of the pipeline using noise data collected in different environments.

## 2.1    Bilateral extension

The approach adopted in (Gopalakrishna, et al. 2012) is that the noise suppression structure is different for each noisy environment. Then, by detecting a noise class, the system switches to the appropriate structure. Now, the approach introduced in this dissertation is that the environment-dependent structure is the case not only for the noise suppression but also for different directions. One realization of this extension is depicted in Figure 2.1 by considering the parameter space consisting of noise suppression parameters plus directional parameters. The bilateral extension using only a single processor is achieved based on a gain model considered to be a function of both Signal to Noise Ratios (SNRs) and source angles (directions). In other words, the gain function used in the unilateral pipelines (Erkelens, Jensen and Heusdens 2007; Gopalakrishna, et







al. 2012) becomes different along different directions in the bilateral extension. Its basic enhancement part is illustrated in Figure 2.2. As it will be discussed later, this approach leads to an effective pipeline for bilateral speech processing with respect to computations, memory requirements and hardware efficiency. By performing the extension using only a single processor, the synchronization issues regarding the binaural stimulation are resolved.

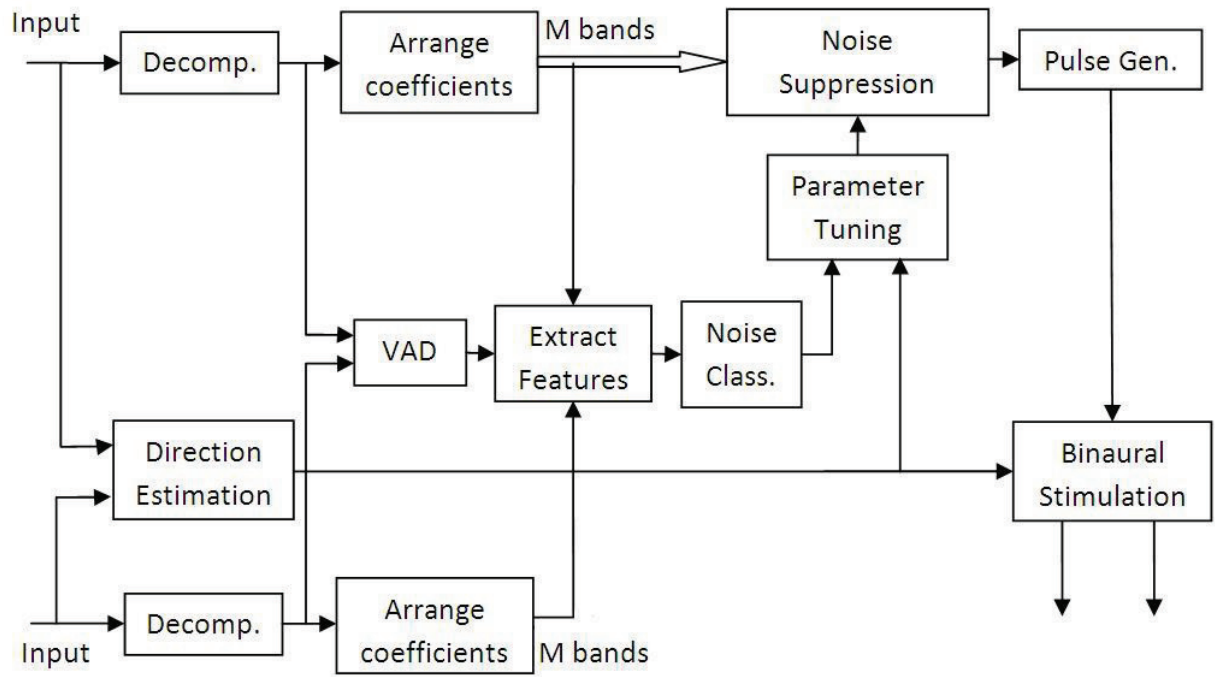

Figure 2.1. Block diagram representation of the extension to bilateral cochlear implant pipelines.

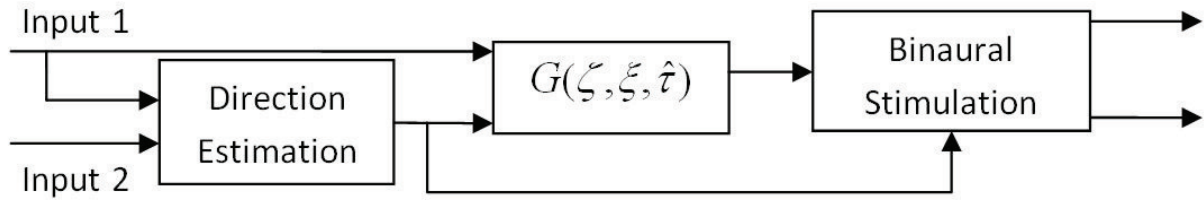

Figure 2.2. Gain function based on the developed suppression scheme



Basically, gain $G$ is applied only onto the first input, though it is not only a function of estimated prior SNR $(\zeta)$ and posterior SNR $(\xi)$ but also dependent on the time delay estimation between the two inputs. This time delay estimate is the Interaural Time Difference (ITD) in binaural hearing. It is denoted by $\hat{\tau}$ here, as shown in Figure 2.2. In addition to the third argument of the gain function, the second input is used as extra information to increase the reliability of the Voice Activity Detector (VAD) and the noise classification components. All the processing in the main path including the noise suppression is performed on the first input only. Stimulation signals representing the second input is reconstructed using the directional information estimated from the ITD algorithm.

Although this approach is computationally attractive, it might be prone to non-negligible distortions. This drawback becomes more pronounced if the $\hat{\tau}$ estimate is not precise enough. To address this issue, the optimization of the direction-dependent gain parameters is modeled with respect to the distortion of not only the first input but also the reconstructed second input. In this manner, the distortion measure is defined to minimize Mean Square Error (MSE) for both the inputs.

Suppose $X_1$ and $\hat{X}_1$ are the clean and enhanced signals for input 1 and $D(X_1, \hat{X}_1)$ is the distortion measure used to optimize the gain parameters, then

$$\tilde{G}_{\blacksquare} = arg \underset{G_{ij}}{\blacksquare} in \, D(X_1, \hat{X}_1) \qquad (2.1)$$

where $G_{ij}$ denotes the gain parameter corresponding to the cell representing the $i$-th partition of the prior SNR range and the $j$-th partition of the posterior SNR range. If the range of prior and



posterior SNR estimates are partitioned to a total of $I$ and $J$ values, respectively, the gain is an $I \times J$ matrix containing the noise suppression parameters, that is

$$\mathbf{G} = \{G_{ij}, \forall i = 1, \ldots, I, \forall j = 1, \ldots, J\} \qquad (2.2)$$

Now assume that the following directionality distortion measure incorporating both of the inputs is minimized, that is

$$\tilde{G}_{\blacksquare} = arg\,\blacksquare \min_{G_{ijl}} \{D(X_1, \hat{X}_1) + D(X_2, \hat{X}_2)\} \qquad (2.3)$$

where $G_{ijl}$ is the gain parameter corresponding to the cell representing the $i$-th partition of the prior SNR range and the $j$-th partition of the posterior SNR range and the $l$-th partition of the time delay range. Different weights could be assigned to the two distortion measures as stated in the next section. This is exhibited in Figure 2.3, where $Y_1$ and $Y_2$ in this figure denote the original noisy inputs.

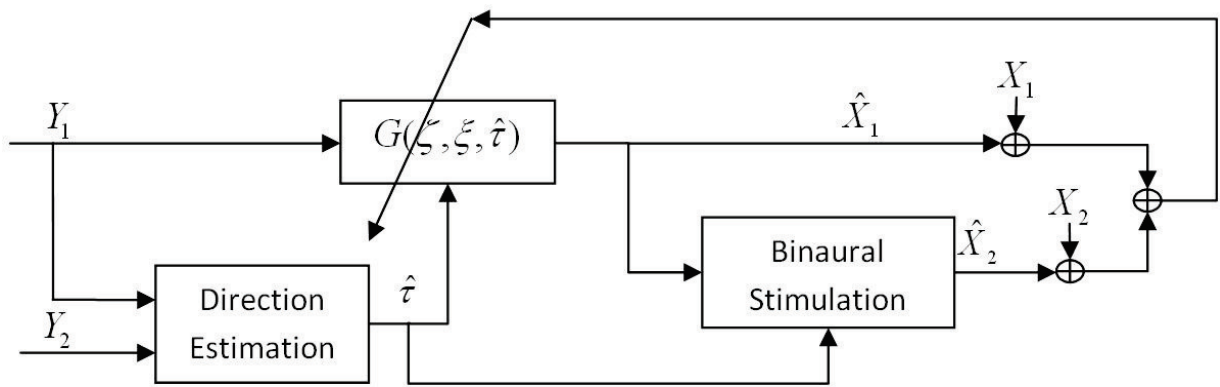

Figure 2.3. Optimization process of data-driven noise suppression gain.



### 2.1.1  Euclidean distortion measure

As outlined in Section 2.1, noise suppression is applied to a reference signal according to the prior and posterior SNR estimates. The reference signal is defined as the input which arrives first from the right or left microphone and the other one is defined as the non-reference signal. All the associated variables are distinguished by the subscripts $r$ and $nr$, respectively. In other words, the reference signal is the first input when the delay is positive, and it is the second one if it is negative.

After applying the suppression gain on the reference signal, an enhanced reference signal is obtained. This gain is a function of prior and posterior SNRs.

Next, the enhanced non-reference signal is obtained by applying another gain $\mathbf{H}$ which is a function of the time difference of the estimate of the direction of arrival,

$$\mathbf{H} = \{H_l, \forall l = 1, \ldots, L\} \tag{2.4}$$

This constitutes a simple representation of the Head-Related Transfer Function (HRTF) where $H_l$ values are assumed to be only a function of delay. This function is a representation of the Interaural Level Difference (ILD). The ITD and ILD are the two main binaural cues for sound localization that are modeled in this representation. HRTF is also assumed to be the same for both left-to-right and right-to-left directions. As a result, the problem of identifying these transfer functions is transformed to the problem of estimating the time difference of arrival, thereby utilizing a data-driven approach to identify ILDs. There are many methods developed for this purpose in the literature; here the Generalized Cross Correlation technique is utilized as it provides an appropriate compromise between accuracy and computational complexity (Chen, Benesty and Huang 2006).



The objective is to find all $G_{ij}$ and $H_l$ parameters by optimizing a distortion measure defined based on both the reference and non-reference outputs. Having estimated the prior and posterior SNRs and the time delay in a given fame, the corresponding cells of $G_{ij}$ and $H_l$ can then be determined. Because $G_{ij}$'s are applied only onto the reference input, their optimal values are independent of the non-reference signal. To find optimal $G_{ij}$'s via the Minimum Mean Square Error (MMSE) minimization, noisy and clean data points of the reference signal need to be stored during an offline data collection process.

If $R_{r,ij}(m)$ is the $m$-th data-point of the magnitude spectrum of the noisy reference signal and $A_{r,ij}(m)$ is the corresponding clean signal for which the estimated prior and posterior SNRs fall into the $(i, j)$-th cell of the table $\mathbf{G}$, this dataset is needed to be stored for the optimization of $\mathbf{G}$: $\{R_{r,ij}(m)\}_{m=1}^{M_{ij}}, \{A_{r,ij}(m)\}_{m=1}^{M_{ij}}$, where $M_{ij}$ denotes the total number of data points observed at the SNRs corresponding to $(i, j)$. Note that $m$ is only an index for pairs of noisy amplitude $R_{r,ij}(m)$ and corresponding clean amplitude $A_{r,ij}(m)$ which fall into the $(i, j)$-th cell of $\mathbf{G}$. Then, for $M_{ij}$ number of train data points for each $(i, j)$-th parameter of $\mathbf{G}$ or $G_{ij}$, a total of $\sum_{i=1}^{I} \sum_{j=1}^{J} M_{ij}$ data points need to be collected from the reference signal for training purposes.

The gain $\mathbf{H}$ is applied onto the enhanced reference signal to obtain the enhanced non-reference output. Therefore, the corresponding clean non-reference output spectrum is stored as $\{R_{r,i'j'l}(m')\}_{m'=1}^{M'_{ijl}}, \{A_{nr,ijl}(m')\}_{m'=1}^{M'_{ijl}}$, where $(R_{r,i'j'l}(m'), A_{nr,ijl}(m'))$ denotes the $m'$-th pair of data for the $(i, j)$-th estimate of the SNRs and when $\hat{\tau}$ falls into the $l$-th cell of the vector $\mathbf{H}$, and $M'_{ijl}$



denotes the total number of observations in the $(i, j)$-th estimate of SNRs and $l$-th estimate of the delay.

The total distortion is considered to be a linear combination of distortions associated with the reference and the non-reference spectral errors, that is

$$D = D_r + \beta D_{nr}, \ \ 0 \le \beta \le 1 \tag{2.5}$$

The parameter $0 \le \beta \le 1$ corresponds to the weight assigned to the error in the non-reference signal versus the reference signal. It is a user-defined parameter which reflects the relative importance of optimization on the non-reference path versus the reference one. A weight of zero corresponds to the regular one-channel (unilateral) case (Erkelens, Jensen and Heusdens 2007) and the non-reference speech signal is left unprocessed. On the other hand, a weight of unity treats the two distortion paths equally. The unilateral case ($\beta = 0$) was shown in a previous work to outperform the fixed suppression approach in terms of a number of quality measures. With a nonzero value of $\beta$, one would deal with a more complicated gain optimization problem as it becomes a joint optimization with an objective function defined over both left and right spectra. Then, a weight of $\beta = 1$ would correspond to the most complicated form of this optimization problem. In the experiments reported in Chapter 6, the worst case results are reported with $\beta$ set to 1.

The distortions $D_r$ and $D_{nr}$ are

$$D_r \equiv \sum_{i=1}^{I} \sum_{j=1}^{J} D_{r,ij} \tag{2.6}$$

with



$$D_{r,ij} \equiv \sum_{m=1}^{M_{ij}} A_{r,ij}^p(m)\{A_{r,ij}(m) - G_{ij}R_{r,ij}(m)\}^2 \qquad (2.7)$$

and

$$D_{nr} \equiv \sum_{i=1}^{I}\sum_{j=1}^{J}\sum_{l=1}^{L} D_{nr,ijl} \qquad (2.8)$$

with

$$D_{nr,ijl} \equiv \sum_{m'=1}^{M_{ijl}'} A_{nr,ijl}^p(m')\{A_{nr,ijl}(m') - G_{ij}H_l R_{r,ijl}(m')\}^2 \qquad (2.9)$$

where $D_{r,ij}$ and $D_{nr,ijl}$ in (2.7) and (2.9) are defined based on the weighted-Euclidean distortion

measure on spectral errors (Loizou 2005; Erkelens, Jensen and Heusdens 2007) and $p$ is a

weighting parameter. A weight of $p = 0$ reduces the problem to the traditional MMSE

optimization problem, while nonzero weights amplify the effect of smaller or larger amplitudes.

For example, a weight of $p = -1$ gives more weight to smaller amplitudes. An analytical study

of the effect of different weights on the derived gain functions and their relations to MMSE and

log-MMSE solutions is provided in (P. Loizou 2005). In the experiments reported in Chapter 6,

the MMSE criterion ( $p = 0$ ) is considered, while the derivations are stated for any weights.

Hence,

$$D = \sum_{i=1}^{I}\sum_{j=1}^{J}\left\{ D_{r,ij} + \beta\sum_{l=1}^{L} D_{nr,ijl} \right\} \equiv \sum_{i=1}^{I}\sum_{j=1}^{J} D_{ij} \qquad (2.10)$$

To find the solution for this optimization problem, let us define the following quantities

$$S_{r,ij,1} \equiv \sum_{m=1}^{M_{ij}} A_{r,ij}^{p+1}(m).R_{r,ij}(m) \qquad (2.11)$$



$$S_{r,ij,2} \equiv \sum_{m=1}^{M_{ij}} A_{r,ij}^p \left( m \right).R_{r,ij}^2 (m) \tag{2.12}$$

$$S_{nr,ijl,1} \equiv \sum_{m'=1}^{M_{ijl}'} A_{nr,ijl}^{p+1} \left( m' \right).R_{r,ijl} (m') \tag{2.13}$$

$$S_{nr,ijl,2} \equiv \sum_{m'=1}^{M_{ijl}'} A_{nr,ijl}^{p} \left( m' \right).R_{r,ijl}^2 (m') \tag{2.14}$$

As a result of setting the partial derivatives to zero, that is

$$\frac{\partial D}{\partial G_{ij}} = \frac{\partial D_{ij}}{\partial G_{ij}} = \frac{\partial D_{r,ij}}{\partial G_{ij}} + \beta \sum_{l=1}^{L} \frac{\partial D_{nr,ijl}}{\partial G_{ij}} = 0 \tag{2.15}$$

one gets

$$S_{r,ij,1} + \beta \sum_{l=1}^{L} H_l S_{nr,ijl,1} = G_{ij} \left[ S_{r,ij,2} + \beta \sum_{l=1}^{L} H_l^2 S_{nr,ijl,2} \right] \tag{2.16}$$

As a result of

$$\frac{\partial D}{\partial H_l} = \sum_{i=1}^{I} \sum_{j=1}^{J} \beta \frac{\partial D_{nr,ijl}}{\partial H_l} = 0 \tag{2.17}$$

one gets

$$\sum_{i=1}^{I} \sum_{j=1}^{J} G_{ij} S_{nr,ijl,1} = H_l \sum_{i=1}^{I} \sum_{j=1}^{J} G_{ij}^2 S_{nr,ijl,2} \tag{2.18}$$

From the above equations, it can be seen that the optimal value of each $G_{ij}$ and $H_l$ depends on

the knowledge of the true value of the other one. A simple approach would be to recursively

update the solution by starting with an appropriate initialization. Here, $G_{ij}$ is initialized with the



weighted-Euclidean response of the regular case (Loizou 2005; Erkelens, Jensen and Heusdens 2007) and the problem is solved in a quasi-static way as follows

$$G_{ij}^{(0)} = \frac{S_{r,ij,1}}{S_{r,ij,2}} \qquad (2.19)$$

Then, the solution is updated according to (2.16) and (2.18) at any iteration step $q$ as follows

$$H_l^{(q)} = \sum_{ij} G_{ij}^{(q-1)} S_{nr,ijl,1} / \sum_{ij} (G_{ij}^{(q-1)})^2 S_{nr,ijl,2}, \forall l = 1,\ldots,L \qquad (2.20)$$

$$G_{ij}^{(q)} = \{S_{r,ij,1} + \beta \sum_l H_l^{(q)} S_{nr,ijl,1}\} / \{S_{r,ij,2} + \beta \sum_l \left(H_l^{(q)}\right)^2 S_{nr,ijl,2}\}, \forall i = 1,\ldots,I, \forall j = 1,\ldots,J \qquad (2.21)$$

This recursion is repeated until a satisfactory convergence is reached. Note that although unlike the unilateral case the optimal values of the gain parameters in (2.16) are not reached in a closed form, it is seen that the optimization problem defined in (2.10) is a convex function of $\mathbf{G}$ and $\mathbf{H}$ gain parameters. Therefore, a sufficient number of recursions stated in (2.20) and (2.21) guarantees reaching the global optimum. In practice, reaching only a close estimation of the global optimal point may be satisfactory and then going through a limited number of recursions can reduce the training time. Initializing the recursion with an appropriate starting gain value is important in this regard. The closed form solution of the unilateral case would serve as a good initialization choice and is used here in (2.19). Obviously, as these computations are all performed offline, attempting to reach a more accurate estimate by considering more number of recursions would not have any effect during the actual operation.



**2.1.2 Components of the single-processor bilateral speech processing pipeline**

To further clarify the entire process, the environment-specific unilateral CI pipeline in (Gopalakrishna, et al. 2012) —shown in Figure 1.1— is outlined below. The operations written inside parentheses are added to the one-channel pipeline for the bilateral extension shown in Figure 2.1.

In each frame:

(1- Estimate time delay between input 1 and input 2.)

2- Decompose input 1 (and input 2) using Wavelet Packet Transform.

3- Use wavelet coefficients to compute the subband power difference for the VAD to determine whether speech data is noise-only.

(4- Combine the VAD outputs from both decompositions to make the final decision on noise-only detection.)

5- If noise is detected by the VAD, extract features using the wavelet coefficients of the decomposition 1 (and decomposition 2).

6- Classify the noise using features extracted for the input 1 (and input 2) or (combine classification results using features of input 1 with those of input 2).

7- If change in the noise class is detected, load the gain function optimized for the corresponding noise environment (and the estimated direction).

8- Estimate prior and posterior SNRs.

9- Apply appropriate gains to the input 1 frequency bands.

10- Extract envelope, select max amplitude channels, compress, and pulse modulate.



(11- Reconstruct input 2 stimulation signal based on input 1 stimulation signal using the estimated time delay.)

With the gain function receiving three inputs of SNRs and direction, the memory requirement remains manageable. It should be noted that the optimization solution for the noise environment is challenging even for a simple distortion. Therefore, here it is considered that the gain on the first input is dependent on SNR estimations, and the second input is dependent on the direction estimation. This is based on the assumption that SNR-dependent and direction-dependent gains are independent. However, still the gain optimization is performed based on the joint distortion of the sources.

Let the reference signal be the input that is recognized based on the estimation of the time difference of arrival. Then, the SNR-dependent gains are applied onto the reference signal and the non-reference enhanced signal is reconstructed by using the direction-dependent gains applied onto the enhanced version of the reference signal. For each noise type, tuning of the gains is performed based on an extension of the data-driven approach in (Erkelens and Heusdens 2008; Ephraim and Malah 1984). The decision-directed approach in (Ephraim and Malah 1984) is used to estimate the prior SNR and the IMCRA (Improved Minima Controlled Recursive Averaging) method (Cohen 2003) is used to estimate noise.

After collecting appropriate training data corresponding to the recognized grid cell, they are used in the MSE optimization of the gain parameters. It is to be noted that here only the solution for the weighted Euclidean distortion measure is stated for the two signals with the weighting parameter $p$ as defined in (2.7) and (2.9). After learning the gain parameters for each



environment, the automatic environment-adaptive suppression can be performed according to the model considered in the above offline gain optimization procedure.

### 2.1.3 Memory requirements

Running the developed bilateral extension requires storing a suppression gain function $\mathbf{G}$ and an HRTF table $\mathbf{H}$ for each environment. In this Section, the storage requirement is discussed. If the bilateral speech processing is performed using two independent processors in parallel or by a single processor but in a sequential manner, it would require storing two suppression gain functions for each noise type. This would be equal to $2 \times I \times J$ number of parameters. Note that this way, synchronization would be required and two processors would be performing the bilateral stimulation. To avoid the synchronization problem while providing the bilateral stimulation using only a single processor, one needs to store $L$ independent suppression gain tables for each direction of a noise type. This would require a storage capacity of $L \times I \times J$ number of gain parameters. Using the developed method, this storage is reduced to storing a suppression gain table and an HRTF vector according to the number of directions. This means that the storage is reduced to only $(I \times J + L)$ number of gain parameters. Based on a word length of $W$ bits for each gain parameter, Table 2.1 shows the memory requirements for each method of the bilateral enhancement. Typical memory requirements for a $60 \times 70$ representation of the suppression gain table, considering 13 different directions, using a word length of 16 bits are also listed. It can be seen that the developed method for the bilateral enhancement requires only 8.2285 kB of storage which is only 0.3% more than what is required for storing one suppression gain table for the unilateral enhancement (16.4063/2 = 8.2032 kB). It is worth



mentioning that this also implies the computational effectiveness superiority of the extended single-processor pipeline against bilateral processing in independent channels or using a single processor but in a sequential manner. The reason is that the developed extension does not require any SNR estimations and gain multiplications on the non-reference input.

Table 2.1. Storage requirements of different bilateral enhancement methods and typical needed memory for a 60×70 suppression gain table, a Head-Related Transfer Function (HRTF) of length 13 directions, and a word length of 16 bits.

| *Hardware System Architecture* | **Double-Processor / Single-Processor (Sequential Processing)** | **Single-Processor (Independent Gains for Different Directions)** | **Single-Processor (Proposed)** |
|---|---|---|---|
| Storage Requirements (bits) | $2 \times I \times J \times W$ | $L \times I \times J \times W$ | $(I \times J + L) \times W$ |
| Typical Memory (kB) | $\approx 16.4$ | $\approx 106.6$ | $\approx 8.2$ |

# CHAPTER 3

## GENERALIZATIONS TO NON-EUCLIDEAN DISTORTION CRITERIA[‡]

Because of its mathematical simplicity, Euclidean distortion is the most commonly used objective function in many speech processing applications. However it is well known that human hearing perception models suggest speech distortions that affect speech understanding based on non-Euclidean functions. A generalized framework is developed in this Chapter that allows one to train suppression and head-related transfer function gain tables not only for different noise environments but also for different distortion criteria in the single-processor bilateral speech processing pipeline developed in Chapter 2. This generalization incorporates any differentiable measure with the unilateral data-driven optimization methods becoming its special cases. Specifically, the solutions for three distortion measures of Weighted-Euclidean, Log-Euclidean and Weighted-Cosh are provided.

## 3.1 Generalized non-Euclidean solutions

To generalize the bilateral data-driven framework defined in (2.2) and (2.4) to non-Euclidean distortion measures, let us redefine $D_r$ in the objective function in (2.5) as the mean of distortions associated with different prior and posterior Signal to Noise Ratios (SNRs),

---







$$D_r \equiv \frac{1}{IJ}\sum_{i=1}^{I}\sum_{j=1}^{J}D_{r,ij} \tag{3.1}$$

where $D_{r,ij}$ is the mean distortion over data observed at the $i$-th prior SNR and the $j$-th posterior

SNR. Similarly, let us define non-reference errors as follows

$$D_{nr} \equiv \frac{1}{IJL}\sum_{i=1}^{I}\sum_{j=1}^{J}\sum_{l=1}^{L}D_{nr,ijl} \tag{3.2}$$

where $D_{nr,ijl}$ is the mean distortion over data observed at the $l$-th direction along with the $i$-th

prior SNR and the $j$-th posterior SNR.

The gradient of the total distortion with respect to the suppression and Head-Related Transfer

Function (HRTF) gain parameters can be written as follows:

$$\frac{\partial D}{\partial G_{ij}} = \frac{1}{IJ}\{\frac{\partial D_{r,ij}}{\partial G_{ij}} + \beta \frac{1}{L}\sum_{l=1}^{L}\frac{\partial D_{nr,ijl}}{\partial G_{ij}}\} \tag{3.3}$$

$$\frac{\partial D}{\partial H_l} = \frac{1}{IJL}\beta\sum_{i=1}^{I}\sum_{j=1}^{J}\frac{\partial D_{nr,ijl}}{\partial H_l} \tag{3.4}$$

Different distortion functions can be used to compute $D_{r,ij}$ and $D_{nr,ijl}$ values. Consider $d(A,\hat{A})$ to

be such a function computing the distortion between the clean spectral amplitude $A$ and the

estimated enhanced counterpart $\hat{A}$. The enhanced reference signal is then obtained by mapping

the noisy reference amplitudes via $\mathbf{G}$. Therefore,

$$D_{r,ij} \equiv \frac{1}{M_{ij}}\sum_{m=1}^{M_{ij}}d(A_{r,ij}(m), G_{ij}R_{r,ij}(m)) \tag{3.5}$$

where $A_{r,ij}(m)$ is the $m$-th data sample of the reference clean spectral amplitude observed at the

prior and posterior SNRs corresponding to the $(i,j)$-th cell of the suppression gain table,



$R_{r,ij}(m)$ is its noisy counterpart and $M_{ij}$ is the total number of data samples collected for this cell.

The non-reference clean amplitudes are estimated by mapping the estimated reference amplitudes and using the HRTF gain $\mathbf{H}$. Hence,

$$D_{nr,ijl} \equiv \frac{1}{M_{ijl}^{'}} \sum_{m'=1}^{M_{ijl}^{'}} d(A_{nr,ijl}(m'), G_{ij} H_l R_{r,ijl}(m')) \tag{3.6}$$

where $A_{nr,ijl}(m')$ and $R_{r,ijl}(m')$ are, respectively, the $m'$-th data sample of the non-reference clean and the reference noisy amplitudes corresponding to the $(i,j)$-th cell of the suppression gain table and the $l$-th cell of the HRTF gain table. The total number of data samples is assumed to be $M_{ijl}^{'}$ for each set.

### 3.1.1 Weighted-Euclidean distortion criterion

Weighted-Euclidean (WE) distortion function with weight $p$ is defined as

$$d_{\mathrm{WE}}(A, \hat{A}) \equiv A^p \cdot (A - \hat{A})^2 \tag{3.7}$$

Considering the definitions in (2.11-2.14), and from (3.3-3.4) and based on the definitions in (3.5-3.7), the WE solutions are derived to be

$$\frac{\partial D}{\partial G_{ij}} = -2\frac{1}{IJ}\{\frac{1}{M_{ij}}(S_{r,ij,1} - G_{ij}S_{r,ij,2}) + \beta\frac{1}{L}\sum_{l=1}^{L}\frac{1}{M_{ijl}^{'}}[H_l S_{nr,ijl,1} - H_l^2 G_{ij} S_{nr,ijl,2}]\} \tag{3.8}$$

$$\frac{\partial D}{\partial H_l} = -2\beta\frac{1}{IJL}\sum_{i=1}^{I}\sum_{j=1}^{J}\frac{1}{M_{ijl}^{'}}[G_{ij}S_{nr,ijl,1} - G_{ij}^2 H_l S_{nr,ijl,2}] \tag{3.9}$$



### 3.1.2   Log-Euclidean distortion criterion

Log-Euclidean (LE) distortion is defined as

$$d_{\mathrm{LE}}(A,\hat{A}) \equiv (\log[A] - \log[\hat{A}])^2 \tag{3.10}$$

Similarly, for this distortion measure, let us define the following terms to obtain a simpler

representation,

$$P_{r,ij} \equiv \log[\prod_{m=1}^{M_{ij}} \frac{A_{r,ij}(m)}{G_{ij}R_{r,ij}(m)}] \tag{3.11}$$

$$P_{nr,ijl} \equiv \log[\prod_{m'=1}^{M'_{ijl}} \frac{A_{nr,ijl}(m')}{G_{ij}H_l R_{r,ij}(m')}] \tag{3.12}$$

From (3.3-3.4) and based on the definitions in (3.5-3.6) and (3.10-3.12), the LE solutions are

derived to be

$$\frac{\partial D}{\partial G_{ij}} = -2\frac{1}{IJ}\frac{1}{G_{ij}}\{\frac{1}{M_{ij}}P_{r,ij} + \beta\frac{1}{L}\sum_{l=1}^{L}\frac{1}{M'_{ijl}}P_{nr,ijl}\} \tag{3.13}$$

$$\frac{\partial D}{\partial H_l} = -2\beta\frac{1}{IJL}\frac{1}{H_l}\sum_{i=1}^{I}\sum_{j=1}^{J}\frac{1}{M'_{ijl}}P_{nr,ijl} \tag{3.14}$$

### 3.1.3   Weighted-Cosh distortion criterion

Weighted-Cosh (WC) distortion with weight $p$ is defined as

$$d_{\mathrm{WC}}(A,\hat{A}) \equiv A^p.(A\big/\hat{A} + \hat{A}\big/A - 1) \tag{3.15}$$

Similarly, by defining

$$C_{r,ij,1} \equiv \sum_{m=1}^{M_{ij}} \frac{A_{r,ij}^{p+1}(m)}{R_{r,ij}(m)} \tag{3.16}$$



$$C_{r,ij,2} \equiv \sum_{m=1}^{M_{ij}} A_{r,ij}^{p-1}\left(m\right) R_{r,ij}\left(m\right) \tag{3.17}$$

$$C_{nr,ijl,1} \equiv \sum_{m^{'}=1}^{M_{ijl}^{'}} \frac{A_{nr,ijl}^{p+1}\left(m^{'}\right)}{R_{r,ijl}\left(m^{'}\right)} \tag{3.18}$$

$$C_{nr,ijl,2} \equiv \sum_{m^{'}=1}^{M_{ijl}^{'}} A_{nr,ijl}^{p-1}\left(m^{'}\right) R_{r,ijl}\left(m'\right) \tag{3.19}$$

and from (3.3-3.4) and based on the definitions in (3.5-3.6) and (3.16-3.19), the following WC solutions are derived,

$$\frac{\partial D}{\partial G_{ij}} = \frac{1}{IJ}\{ -\frac{1}{G_{ij}^{2}}\left[ \frac{1}{M_{ij}} C_{r,ij,1} + \beta \frac{1}{L} \sum_{l=1}^{L} \frac{1}{M_{ijl}^{'}} \frac{1}{H_{l}} C_{nr,ijl,1} \right] + \frac{1}{M_{ij}} C_{r,ij,2} + \beta \frac{1}{L} \sum_{l=1}^{L} \frac{1}{M_{ijl}^{'}} H_{l} C_{nr,ijl,2} \} \tag{3.20}$$

$$\frac{\partial D}{\partial H_{l}} = \frac{1}{IJL}\{ -\beta \frac{1}{H_{l}^{2}} \sum_{i=1}^{I} \sum_{j=1}^{J} \frac{1}{G_{ij}} \frac{1}{M_{ijl}^{'}} C_{nr,ijl,1} + \beta \sum_{i=1}^{I} \sum_{j=1}^{J} G_{ij} \frac{1}{M_{ijl}^{'}} C_{nr,ijl,2} \} \tag{3.21}$$

The derived gain optimization solutions for WE, LE, and WC distortion criteria can be used with any gradient-based optimization technique to optimize gain parameters in the bilateral pipeline developed in Section 2.1.

# CHAPTER 4

# UNIFIED OPTIMIZATION FRAMEWORK[§]

This Chapter develops a generalization of the models in Chapter 2 and Chapter 3 using psychoacoustic human perception to model the head-related transfer function with the flexibility of using different distortion criteria. Instead of time-domain cues used in Chapter 2, phase differences are modeled and estimated as binaural cues for localization in the model developed in this Chapter. This bilateral pipeline is designed in such a way that it adds minimal extra memory and computational load compared to the unilateral case, thus making its deployment practical. The generalization is carried out for two main families of speech distortion criteria, namely amplitude-weighted and loudness-weighted distortions. The developed techniques are primarily meant for bilateral CIs, however, they are general purpose in the sense that they can be easily scaled to binaural hearing aids and dual-channel speech enhancement applications. The developed unified framework covers the data-driven gain optimizations in the unilateral speech processing pipeline and the developed single-processor bilateral pipeline as special cases.

## 4.1    Overview of the framework

As discussed in Section 2.1, the data-driven optimization of noise suppression gain functions has been used to correct for errors introduced due to model incompleteness and Signal to Noise Ratio







(SNR) estimation inaccuracies  (Erkelens, Jensen and Heusdens 2007; Erkelens and Heusdens 2008; Gopalakrishna, et al. 2012). In the previous Chapters, this approach was utilized as the framework for environment-adaptive speech processing where each gain function was optimized using data collected in different noisy environments. This allowed suppressing different noise types without being restricted to a previously conditioned speech and noise distribution. It has been shown that the environment-adaptive pipeline results in speech quality improvements over model-based fixed gain functions (Gopalakrishna, et al. 2012; Mirzahasanloo, et al. 2012). The joint optimization of gain functions for binaural speech processing has been shown to provide even further benefits as it exploits a more general framework (Mirzahasanloo, et al. 2013). Using a data-driven approach, our bilateral solution in (Mirzahasanloo, et al. 2013) not only improved speech quality, but also provided a computationally efficient pipeline using a single processor. The use of a single processor allows generating synchronized bilateral stimulation signals towards more effective utilization of bilateral Cochlear Implants (CIs).

The framework in Chapter 2 used two separate gain functions where a suppression gain was applied on a reference input to provide an enhanced reference speech and a Head-Related Transfer Function (HRTF) was used to reconstruct a non-reference output. The suppression gain was a function of prior and posterior SNRs (Loizou 2007; Ephraim and Malah 1985; Ephraim and Malah 1984), while the HRTF gain was assumed to be a function of delay between reference and non-reference inputs (Mirzahasanloo, et al. 2013). The HRTF gain was characterized by a delay estimation which was derived from a Time Difference of Arrival (TDOA) estimation using a generalized cross correlation (Chen, Benesty and Huang 2006). Although this model resulted in a memory-efficient solution, its performance was sensitive to errors in the estimation of TDOA.



Lowpass filtering of estimated values increased the reliability but limited the capability of tracking direction changes (Mirzahasanloo, et al. 2013). Also, the assumed model could not characterize frequency-dependent variations between the two inputs as the HRTF gain was a function of the inputs derived from the time-domain analysis of the delay. Here, an extension of this model is presented by modeling both the suppression and HRTF gain estimations based on the spectral domain information. Note that this approach provides a more general framework and the optimization of the gain parameters discussed in Chapter 2 cannot be directly used. For this reason, the optimization approach developed in Chapter 3 is considered here in order to obtain solutions of this generalized framework which can cope with non-Euclidean distortion criteria that are needed in some speech processing applications (Loizou 2005).

Figure 4.1 presents the generalized bilateral noise suppression framework for environment-adaptive speech processing in bilateral CIs using a single processor (Mirzahasanloo and Kehtarnavaz, 2014).

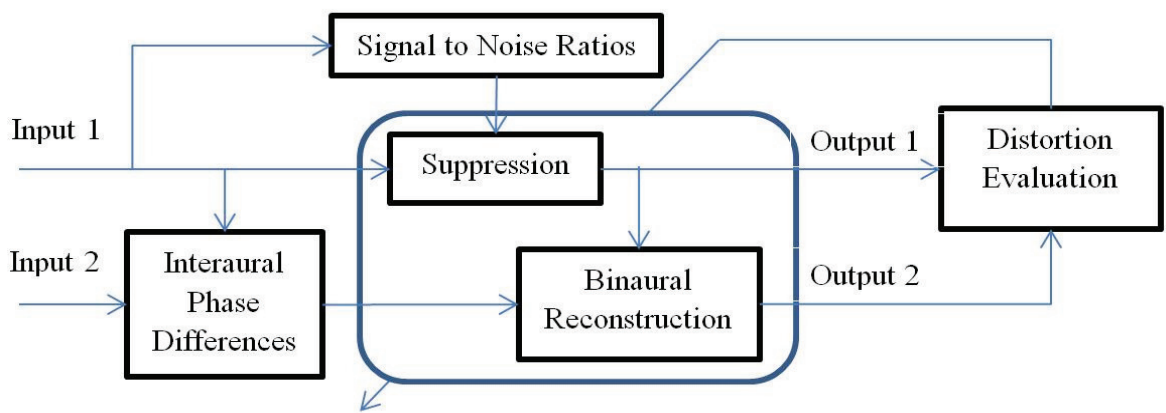

Figure 4.1. Binaural reconstruction and suppression gain optimization architecture of the generalized binaural speech processing framework for bilateral cochlear implants using a single processor.



Input 1 is used to obtain the prior and posterior SNRs and thus to characterize the noise

suppression gain function. One may use any prior SNR and noise power estimation approach

here. Here, the decision-directed approach, which is the most commonly used method in

statistical estimation of noise suppression gain functions (Loizou 2007; Ephraim and Malah

1985), is considered. The Interaural Phase Difference (IPD) information between the two input

signals are used to characterize another function to reconstruct an estimate of the processed

output 2 from the processed output 1, which is provided by the suppression function. The two

estimated outputs are evaluated by a distortion criterion to obtain an appropriate feedback signal

that is used to update currently used suppression and binaural reconstruction gain parameters.

In Section 4.2, the realization of such a model based on a single processor is discussed followed

by the optimization solutions in Section 4.3.

## 4.2    Single-processor approach to bilateral speech enhancement

The entire processing in Figure 4.1 is conducted in the spectral domain which makes it possible

to directly use estimated outputs for the generation of CI stimulation pulses without any extra

processing. The binaural reconstruction component represents estimates of different HRTFs

along different directions. IPD values are computed to provide these directions for the binaural

reconstruction.

Each HRTF along a direction is a function of different frequency bands. These bands are

determined by some non-overlapping partition of the frequency range providing a frequency sub-

banding. Different frequency grouping methods can be used to realize this sub-banding based on

human hearing perception psychoacoustics for audible critical bands (Zwicker 1961). Here, bark



scale is used to form this frequency partitioning. Let $Y_L(\omega)$ and $Y_R(\omega)$ denote the left and right

input noisy spectral amplitudes, then IPDs over different frequency bands are given by

$$\text{IPD}(b) = \angle(\sum_{k=k_b}^{k_{b+1}-1} X_L(\omega_k) X_R(\omega_k) \exp(j(\theta_L(\omega_k) - \theta_R(\omega_k)))), \quad \forall b = 1,...,B, \tag{4.1}$$

where $\omega_{k_b}$ is the start frequency band of subband $b$ and $\theta$ represents phase. These IPD values

are further uniformly discretized into $Q$ different directions to characterize the binaural

reconstruction function $H$ which can be represented by a matrix $\mathbf{H}$ of $Q \times B$ entries

$$\mathbf{H} = \left\{ H_{qb}, \quad \forall q = 1,...,Q, \quad \forall b = 1,...,B \right\}. \tag{4.2}$$

Similarly, the ranges of the prior and posterior SNR estimates are discretized uniformly to

characterize the suppression gain function. If $I$ different prior SNR and $J$ different posterior

SNR partitions are considered,

$$\mathbf{G} = \left\{ G_{ij}, \quad \forall i = 1,...,I, \quad \forall j = 1,...,J \right\}. \tag{4.3}$$

The availability of the gain parameters in (4.2-4.3) allows the binaural processing pipeline in

Figure 4.1 to provide binaural outputs using a single processor.

Compared to (Mirzahasanloo, et al. 2013) and the model developed in Chapter 2, an important

advantage of this extension to band-specific frequency-dependent HRTF estimation is that since

all the processing is brought into the frequency domain, there is no need for any time-domain

TDOA estimation as done in (Mirzahasanloo, et al. 2013). As a result, inaccuracies in the delay

estimation do not adversely affect the frequency-domain optimization. In (Mirzahasanloo, et al.

2013), an attempt was made to decrease the effects of such delay estimation inconsistencies by

median filtering of the TDOA estimated sequence. Although this approach provided an



acceptable performance, it limited the tracking capability in response to rapid (fast) directional changes. Therefore, a tradeoff between lower distortions due to TDOA estimation inaccuracies and the direction tracking capability should be established for not only parameter tuning but also for finding a general parameter set that work optimally in all situations. Furthermore, the delay estimation to compute TDOA usually involves computationally intensive cross-correlation operations.

With the developed extension, there is no need to decide which of the input signals is a reference signal, as long as the range of phase discretization is considered to cover both negative and positive phases, that is from 0 to $2\pi$. However, in order to save memory, if a symmetric HRTF with respect to phase is assumed, then it is required to determine which one of the left and right signals is the reference one. In the introduced framework shown in Figure 4.1, the right and left channels are interchangeable without loss of generality.

The gain table $\mathbf{H}$ can be trained using the left and right data that are generated by considering different binaural cues including monaural head-shadow cues, interaural differences in time (ITD) and level (ILD). In Chapter 2, ITD and ILD equivalent representations were modeled for binaural hearing. Here, instead of time-domain cues, phase differences are modeled and estimated as binaural cues for localization.

## 4.3    Gain estimation

Gain solutions when minimizing different distortions will be different. Because the developed framework does not rely on closed-form solutions for gain parameters, any differentiable distortion criterion can be considered. Three commonly used distortion criteria in speech quality assessment applications are considered here, and the corresponding optimization solutions are



found. Features and characteristics of each criterion are discussed and guidelines are given regarding which criterion to choose depending on the application or noise environments. Essentially, the problem of characterizing an optimal single-processor binaural speech processing pipeline in Figure 4.1 has been reduced to finding gain parameters defined in (4.2-4.3). The objective is to minimize a distortion function defined on both of the output estimations from these gain parameters. The distortion is defined as a weighted linear combination of the distortions on the two outputs. A fixed weighting of the two output distortions is assumed here, but this parameter can be made adaptive in general. If $D_L$ and $D_R$ denote the left and right distortions, respectively, the total distortion can be expressed as

$$D = D_L + \beta D_R, \quad 0 \leq \beta \leq 1. \tag{4.4}$$

Each of the distortions are assumed to be the average of the distortions observed in each pair of the prior and posterior SNR estimates, that is

$$D_L = \frac{1}{IJ} \sum_{i=1}^{I} \sum_{j=1}^{J} D_{L,ij}, \tag{4.5}$$

$$D_R = \frac{1}{IJ} \sum_{i=1}^{I} \sum_{j=1}^{J} D_{R,ij}, \tag{4.6}$$

where the distortions in (4.6) over outputs provided by the binaural reconstruction are also affected by the distortions in each specific direction and frequency band. These distortions are averaged as follows:

$$D_{R,ij} = \frac{1}{QB} \sum_{q=1}^{Q} \sum_{b=1}^{B} D_{R,ijqb}, \quad \forall i = 1, \ldots, I, \quad \forall j = 1, \ldots, J. \tag{4.7}$$



By incorporating the relations in (4.5-4.7), the partial derivatives of the distortion defined in (4.4)

with respect to each suppression gain parameter in (4.3) and each HRTF gain parameter in (4.2),

can be written as

$$\frac{\partial D}{\partial G_{ij}} = \frac{1}{IJ}\left\{\frac{\partial D_{L,ij}}{\partial G_{ij}} + \beta\frac{1}{QB}\sum_{q=1}^{Q}\sum_{b=1}^{B}\frac{\partial D_{R,ijqb}}{\partial G_{ij}}\right\}, \quad \forall i = 1,...,I, \quad \forall j = 1,...,J, \tag{4.8}$$

$$\frac{\partial D}{\partial H_{qb}} = \beta\frac{1}{IJ}\sum_{i=1}^{I}\sum_{j=1}^{J}\frac{1}{QB}\frac{\partial D_{R,ijqb}}{\partial H_{qb}}, \quad \forall q = 1,...,Q, \quad \forall b = 1,...,B. \tag{4.9}$$

As it can be seen, the three terms $\dfrac{\partial D_{L,ij}}{\partial G_{ij}}$, $\dfrac{\partial D_{R,ijqb}}{\partial G_{ij}}$ and $\dfrac{\partial D_{R,ijqb}}{\partial H_{qb}}$ need to be computed for the

solutions in (4.8-4.9). These terms can be represented as a linear combination of data-dependent

and parameter-dependent quantities which, in general, may be non-linear functions of data and

parameters.

In the unilateral suppression model in (Gopalakrishna, et al. 2010; Erkelens, Jensen and

Heusdens 2007) involving the lookup table representation of gains, the parameter-dependent

quantities are linear functions of the gain parameters, thus a closed-form solution can be

obtained. A general model is considered here where both data and parameter quantities can be

nonlinear functions with no closed-form solution for the distortion optimization problem.

Distortion functions are usually defined as some form of dissimilarity measure between the

actual output and the model-estimated output. The gradient of the distortion with respect to the

model parameters is thus taken as a combination of the model gradient weighted by the actual

and estimated outputs. Consequently, the distortion gradients become a combination of the two

sets of data and parameter quantities as defined below



$$\frac{\partial D_{L,ij}}{\partial G_{ij}} = K_{L,ij}^{0} + K_{L,ij}^{\Phi}\Phi_{L,ij} + K_{L,ij}^{\Psi}\Psi_{L,ij}, \tag{4.10}$$

$$\frac{\partial D_{R,ijqb}}{\partial G_{ij}} = K_{R,ijqb}^{0} + K_{R,ijqb}^{\Phi}\Phi_{R,ijqb} + K_{R,ijqb}^{\Psi}\Psi_{R,ijqb}, \tag{4.11}$$

$$\frac{\partial D_{R,ijqb}}{\partial H_{qb}} = \Lambda_{R,ijqb}^{0} + \Lambda_{R,ijqb}^{\Phi}\Phi_{R,ijqb} + \Lambda_{R,ijqb}^{\Psi}\Psi_{R,ijqb}, \tag{4.12}$$

where $K$ and $\Lambda$ are parameter-dependent, and $\Phi$ & $\Psi$ are data-dependent quantities.

Then, the problem of finding the distortion gradient solution reduces to the problem of finding $K$, $\Lambda$, $\Phi$, and $\Psi$.

These characterizing quantities for the two classes of distortions namely amplitude-weighted and loudness-weighted distortions are found next. The solutions for two measures from the first class and one measure from the second class are considered noting that these three distortion measures are most commonly used in speech processing applications.

### 4.3.1    Amplitude-weighted distortions

Amplitude-weighted distortions cover a large set of measures weighted by clean spectral amplitudes given a dissimilarity function between the clean amplitudes $A$ and enhanced spectral amplitudes $\hat{A}$, which can be expressed as

$$d^{w}(A, \hat{A}) = A^{p}.d(A, \hat{A}), \tag{4.13}$$

where $p$ denotes a weighting parameter and $d$ an underlying dissimilarity function.

For this class of distortion measures, the data-dependent quantities in the model are defined in (4.10-4.12) to be some distortion-specific quantities weighted by amplitudes with the corresponding $p$ parameter as follows:



$$\Phi_{L,ij} \triangleq \frac{1}{M_{ij}} \sum_{m=1}^{M_{ij}} A_{L,ij}^p(m) \varphi_{L,ij}(m), \tag{4.14}$$

$$\Psi_{L,ij} \triangleq \frac{1}{M_{ij}} \sum_{m=1}^{M_{ij}} A_{L,ij}^p(m) \psi_{L,ij}(m), \tag{4.15}$$

$$\Phi_{R,ijqb} \triangleq \frac{1}{M_{ijqb}^{'}} \sum_{m'=1}^{M_{ijqb}^{'}} A_{R,ijqb}^p(m') \varphi_{R,ijqb}(m'), \tag{4.16}$$

$$\Psi_{R,ijqb} \triangleq \frac{1}{M_{ijqb}^{'}} \sum_{m'=1}^{M_{ijqb}^{'}} A_{R,ijqb}^p(m') \psi_{R,ijqb}(m'). \tag{4.17}$$

The basis functions $\varphi_L$, $\psi_L$, $\varphi_R$ and $\psi_R$ that are weighted by the clean spectral amplitudes in (4.14-4.17) represent the dissimilarity in the optimization and are different for different distortion measures. In what follows, these basis functions are derived for the Euclidean and Cosh measures, then the solutions minimizing the weighted-Euclidean (WE) and weighted-Cosh (WC) distortions (Loizou 2007; Loizou 2005; Fingscheidt, Suhadi and Stan 2008; Erkelens, Jensen and Heusdens 2007) are found.

*Weighted-Euclidean*

The Euclidean measure (Loizou 2007; Loizou 2005; Fingscheidt, Suhadi and Stan 2008; Mirzahasanloo and Kehtarnavaz 2013a) is defined as the distance between the actual and estimated clean spectral amplitudes where the weighted distortion in (4.13) is expressed as

$$d^E(A, \hat{A}) = \frac{1}{2}(A - \hat{A})^2. \tag{4.18}$$

By defining $\varphi$ and $\psi$ quantities as $\varphi^E$ and $\psi^E$ for the left signal in the equations below

$$\varphi_{L,ij}^E(m) \triangleq A_{L,ij}(m) R_{L,ij}(m), \quad \forall m = 1, ..., M_{ij}, \tag{4.19}$$



$$\psi_{L,ij}^{E}(m) \triangleq R_{L,ij}^{a}(m), \quad \forall m = 1,...,M_{ij}, \tag{4.20}$$

and similarly for the right signal

$$\varphi_{R,ijqb}^{E}(m') \triangleq A_{R,ijqb}(m')R_{R,ijqb}(m'), \quad \forall m' = 1,...,M_{ijqb}^{'}, \tag{4.21}$$

$$\psi_{R,ijqb}^{E}(m') \triangleq R_{R,ijqb}^{a}(m'), \quad \forall m' = 1,...,M_{ijqb}^{'}, \tag{4.22}$$

the following parameter-dependent quantities can then be used to characterize the gradient

solutions for the WE distortion

$$\begin{array}{lll}
\mathrm{K}_{r,ij}^{0^{WE}} = 0 & \mathrm{K}_{nr,ijlb}^{0^{WE}} = 0 & \Lambda_{nr,ijlb}^{0^{WE}} = 0 \\
\mathrm{K}_{r,ij}^{\Phi^{WE}} = -1, & \mathrm{K}_{nr,ijlb}^{\Phi^{WE}} = -H_{lb}, \text{ and } & \Lambda_{nr,ijlb}^{\Phi^{WE}} = -G_{ij} \\
\mathrm{K}_{r,ij}^{\Psi^{WE}} = G_{ij} & \mathrm{K}_{nr,ijlb}^{\Psi^{WE}} = H_{lb}^{2}G_{ij} & \Lambda_{nr,ijlb}^{\Psi^{WE}} = G_{ij}^{2}H_{lb}
\end{array} \tag{4.23}$$

These parameter-dependent quantities along with the basis functions (4.19-4.22) used in the

weighting procedure in (4.14-4.17) complete the solution required to obtain the gradients in

(4.10-4.12) for the WE distortion measures defined in (4.13) and (4.18).

*Weighted-Cosh*

Weighted-Cosh (WC) distortion (Loizou 2007; Loizou 2005; Fingscheidt, Suhadi and Stan 2008;

Mirzahasanloo and Kehtarnavaz 2013a) is another member of the family of amplitude-weighted

distortions which based on (4.13) can be expressed as

$$d^{C}(A,\hat{A}) = (\frac{A}{\hat{A}} + \frac{\hat{A}}{A} - 1). \tag{4.24}$$

Again, by defining $\varphi$ and $\psi$ quantities as $\varphi^{C}$ and $\psi^{C}$ for the left signal in the equations below

$$\varphi_{L,ij}^{C}(m) \triangleq \frac{A_{L,ij}(m)}{R_{L,ij}(m)}, \quad \forall m = 1,...,M_{ij}, \tag{4.25}$$



$$\psi_{L,ij}^C(m) \blacksquare \frac{R_{\ldots}(m)}{A_{L,ij}(m)}, \quad \forall m = 1, \ldots, M_{ij}, \tag{4.26}$$

and similarly for the right channel

$$\varphi_{R,ijqb}^C(m') \blacksquare \frac{A_{\ldots ijqb}(m')}{R_{R,ijqb}(m')}, \quad \forall m' = 1, \ldots, M_{ijqb}', \tag{4.27}$$

$$\psi_{R,ijqb}^C(m') \blacksquare \frac{R_{\ldots ijqb}(m')}{A_{R,ijqb}(m')}, \quad \forall m' = 1, \ldots, M_{ijqb}', \tag{4.28}$$

the following parameter-dependent quantities can then be used to characterize the gradient solutions for the WC distortion

$$
\begin{aligned}
\mathrm{K}_{r,ij}^{0^{WC}} &= 0 & \mathrm{K}_{nr,ijlb}^{0^{WC}} &= 0 & \Lambda_{nr,ijlb}^{0^{WC}} &= 0 \\
\mathrm{K}_{r,ij}^{\Phi^{WC}} &= -\frac{1}{G_{ij}^2}, & \mathrm{K}_{nr,ijlb}^{\Phi^{WC}} &= -\frac{1}{G_{ij}^2}\frac{1}{H_{lb}}, \text{ and } & \Lambda_{nr,ijlb}^{\Phi^{WC}} &= -\frac{1}{H_{lb}^2}\frac{1}{G_{ij}} \\
\mathrm{K}_{r,ij}^{\Psi^{WC}} &= 1 & \mathrm{K}_{nr,ijlb}^{\Psi^{WC}} &= H_{lb} & \Lambda_{nr,ijlb}^{\Psi^{WC}} &= G_{ij}
\end{aligned}
\tag{4.29}
$$

Using these parameter-dependent quantities and data-dependent quantities based on the functions (4.25-4.28) in the general models of (4.14-4.17), the gradient solutions of (4.10-4.12) can be determined which provide the WC-based minimization of (4.4).

### 4.3.2 Loudness-weighted distortions

In loudness-weighted distortions (Loizou 2007; Loizou 2005; Fingscheidt, Suhadi and Stan 2008; Mirzahasanloo and Kehtarnavaz 2013a), in contrast to the amplitude-weighted ones, instead of multiplying the functions $\varphi$ and $\psi$ with spectral amplitudes, they are transferred to the log spectral domain as follows

$$\Phi_{L,ij} \triangleq \frac{1}{M_{ij}}\sum_{m=1}^{M_{ij}}\log[\varphi_{L,ij}(m)], \tag{4.30}$$



$$\Psi_{L,ij} \triangleq \frac{1}{M_{ij}} \sum_{m=1}^{M_{ij}} \log[\psi_{L,ij}(m)], \tag{4.31}$$

$$\Phi_{R,ijqb} \triangleq \frac{1}{M'_{ijqb}} \sum_{m'=1}^{M'_{ijqb}} \log[\varphi_{R,ijqb}(m')], \tag{4.32}$$

$$\Psi_{R,ijqb} \triangleq \frac{1}{M'_{ijqb}} \sum_{m'=1}^{M'_{ijqb}} \log[\psi_{R,ijqb}(m')]. \tag{4.33}$$

By considering different basis functions of $\varphi$ and $\psi$ as dissimilarities in (4.30-4.33), a wide range of loudness-weighted distortions can be defined. Here, the most commonly used distortion criterion of log-Euclidean is considered which is a special case when basis dissimilarities are defined as the Euclidean distance.

### Log-Euclidean

The Euclidean distance in the log-spectral domain is defined as

$$d^{LE}(A,\hat{A}) = \frac{1}{2}(\log[A] - \log[\hat{A}])^2. \tag{4.34}$$

By defining $\varphi$ and $\psi$ quantities as $\varphi^{LE}$ and $\psi^{LE}$ for the left signal in the equations below

$$\varphi_{L,ij}^{LE}(m) \triangleq A_{L,ij}(m), \quad \forall m = 1,...,M_{ij}, \tag{4.35}$$

$$\psi_{L,ij}^{LE}(m) \triangleq R_{L,ij}(m), \quad \forall m = 1,...,M_{ij}, \tag{4.36}$$

and similarly for the right signal

$$\varphi_{R,ijqb}^{LE}(m') \triangleq A_{R,ijqb}(m'), \quad \forall m' = 1,...,M'_{ijqb}, \tag{4.37}$$

$$\psi_{R,ijqb}^{LE}(m') \triangleq R_{R,ijqb}(m'), \quad \forall m' = 1,...,M'_{ijqb}, \tag{4.38}$$



the following parameter-dependent quantities can then be used to characterize the gradient

solutions for the LE distortion

$$
\begin{aligned}
&\mathrm{K}_{r,ij}^{0^{LE}} = \frac{1}{G_{ij}}\log[G_{ij}] \quad \mathrm{K}_{nr,ijlb}^{0^{LE}} = \frac{1}{G_{ij}}\log[G_{ij}H_{lb}] \qquad \Lambda_{nr,ijlb}^{0^{LE}} = \frac{1}{H_{lb}}\log[G_{ij}H_{lb}] \\
&\mathrm{K}_{r,ij}^{\Phi^{LE}} = -\frac{1}{G_{ij}} \qquad , \ \mathrm{K}_{nr,ijlb}^{\Phi^{LE}} = -\frac{1}{G_{ij}} \qquad , \text{and} \ \Lambda_{nr,ijlb}^{\Phi^{LE}} = -\frac{1}{H_{lb}} \\
&\mathrm{K}_{r,ij}^{\Psi^{LE}} = \frac{1}{G_{ij}} \qquad \mathrm{K}_{nr,ijlb}^{\Psi^{LE}} = \frac{1}{G_{ij}} \qquad \Lambda_{nr,ijlb}^{\Psi^{LE}} = \frac{1}{H_{lb}}
\end{aligned}
\tag{4.39}
$$

With these quantities and data-dependent ones in (4.30-4.33) based on (4.35-4.38), the gradient

solutions for (4.4) is completely characterized based on the LE distortion in (4.34).

# CHAPTER 5

# ENVIRONMENT DETECTION IMPROVEMENTS[**]

In this Chapter, the improvements on noise classification accuracy using dual microphones and the addition of quiet and music detection capabilities to the environment-adaptive pipeline are presented. In Section 5.1, a brief overview of the environment detection components as used in Figure 1.1 is provided. An improved noise classification in this pipeline is then presented in Section 5.2. This improvement is achieved by using a dual-microphone and by using a computationally efficient feature-level combination approach. Addition of quiet and music detection capabilities to the pipeline is presented in Section 5.3. A modified Voice Activity Detector is mentioned which provides quiet frame detection in addition to voice and no-voice activity detection in a computationally efficient manner. Music detection is achieved via a two-class Gaussian mixture model classifier requiring no extra computation for feature extraction.

## 5.1    Single microphone noise detection

As shown in Figure 1.1, the previously developed cochlear implant speech processing pipeline consists of two main parallel paths: a speech decomposition path and a noise detection/classification path. As discussed in (Gopalakrishna, Kehtarnavaz and Loizou 2010a), a recursive Wavelet Packet Transform (WPT) is used to decompose the input speech signal into different frequency bands. After appropriately applying a gain function to the magnitude spectrum to suppress noise, channel envelopes are extracted by combining the wavelet packet coefficients of the bands which fall in the frequency range of a particular channel.

---

[**] ©(2013), IEEE. Portions reprinted with permission from (Mirzahasanloo, T., N. Kehtarnavaz, and I. Panahi. "Adding quiet and music detection capabilities to FDA-approved cochlear implant research platform." Proceedings of 8th International Symposium on Image and Signal Processing and Analysis. 2013) and (Mirzahasanloo, T. S., and N. Kehtarnavaz. "Real-time dual-microphone noise classification for environment-adaptive pipelines of cochlear implants." Proceedings of IEEE Int. Conf. on Eng. Med. Biol. 2013)

Followed by rectification, low-pass filtering and envelope compression, stimulation pulses for implanted electrodes are generated using the recursive wavelet decomposition (Gopalakrishna, Kehtarnavaz and Loizou 2010a; Gopalakrishna, Kehtarnavaz and Loizou 2010b). On the other hand, the noise classification path first uses a Voice Activity Detector (VAD) to determine if a current frame is speech or noise. If noise, appropriate noise features are extracted to determine the class or type of the noise environment. The VAD is done using an adaptive threshold for sub-band power which is computed using wavelet coefficients. A wavelet-based VAD is utilized here since the coefficients are already computed as part of the decomposition component, hence making the VAD computationally efficient.

Different noise classifiers along with different features were studied extensively for this pipeline in the previous work in terms of classification performance and computational complexity (Gopalakrishna, et al., 2010). In the final system implementation reported in (Gopalakrishna, et al. 2012), a combination of 13 MFCC (Mel-Frequency Cepstral Coefficients) features with their corresponding first derivatives is used to form a 26-dimensional feature vector. This feature vector is then fed into a Gaussian Mixture Model (GMM) classifier with two clusters. The main attribute of the above pipeline is that it can run in real-time on the FDA-approved PDA research platform for CI studies (Ali, Lobo and Loizou 2012).



## 5.2    Dual microphone noise classification

In this Section, the use of a dual-microphone where two input signals are captured, is considered. A comparison of two approaches using a dual-microphone is made, leading to the selection of the more computationally and memory efficient approach.

The first approach consists of combining decisions given by two classifiers running in parallel each classifying one signal source independently, then using a decision combination module to generate a combined decision outcome. The second approach consists of fusing the feature information extracted from each signal and then using only one classifier.

Decision-level combination can be implemented by training a right and a left GMM classifier independently and combining their decisions within a majority voting strategy. This requires training two independent GMM classifiers and having enough memory space to store two sets of GMM parameters.  Feature-level combination can be implemented by appending the feature vectors to form a single feature vector with twice the dimension. This approach would only require the use of one GMM classifier.

Table 5.1 compares the decision-level and feature-level classification approaches when using a dual-microphone in terms of memory efficiency, computational efficiency, and offline training workload. As the total number of GMM parameters for classifying a (26+26)-dimensional vector is less than that of 2 sets of GMM parameters for classifying a 26-dimensional vector, the feature-level combination requires less memory. The table also shows that the feature-level approach outperforms the decision-level approach in terms of the computation or speed aspect. Another advantage is that the offline training is performed for only one classifier when using the feature-level approach.



Table 5.1. Comparison of feature-level and decision-level classification approaches.

| Comparisons/Approaches | Feature-Level Combination | Decision-Level Combination |
|---|---|---|
| **Memory Efficiency** | 1 Set of GMM parameters for 1 input of 52-dimensional feature vector | 2 Sets of GMM parameters for 2 inputs of 26-dimensional feature vectors |
| **Computational Efficiency** | 1 GMM classification + 1 majority voting | 2 GMM classifications + 2 majority voting + 1 decision combination |
| **Offline Training Workload** | 1 GMM training | 2 GMM training |

Therefore, due to the memory and computational efficiency advantages of the feature-level approach, this approach is adopted in order to improve the classification performance of the environment-adaptive pipelines of CIs.

## 5.3    Quiet and music detection

Adding music and quiet detection capability to the pipeline in Figure 1.1 is performed in such a way that minimal extra computations are added to the pipeline. For the addition of the quiet condition capability, the previously utilized VAD (Gopalakrishna, et al. 2012; Stadtschnitzer, Pham and Chien 2008) is thus modified instead of adding a separate quiet detection component



to the pipeline. A straightforward extension is considered here so that not only voice and noise frames are separated but also quiet frames.

On the other hand, instead of considering music as a separate noise class, a two-class classifier is considered to distinguish music from noise. This way the amount of extra computation is kept quite low. Note that since the same feature vector is used for noise classification, the feature extraction is computed only once for both of the classifiers (music detection and noise category classification). Therefore, no extra computation for feature extraction is introduced for achieving a music detection capability.

Figure 5.1 demonstrates the logical flow of adding these capabilities to the pipeline. When the modified VAD outputs a voice-active frame, no change in the parameters of the suppression algorithm is made. When the frame is indicated to be quiet, no processing is performed on the input speech by turning off the suppression algorithm. When a current frame is detected as neither voice-active nor quiet, the music/noise classifier is activated to determine if it is music or noise. In case of music frames, the noise suppression is turned off or any desired music processing can get activated here. In case of noise frames, the previously developed noise classifier is utilized to determine the noise type in order to appropriately tune the noise suppression parameters.

### 5.3.1 Quiet detection

Now, the details of the VAD extension to quiet condition detection are provided and a measure to quantify its performance is presented.



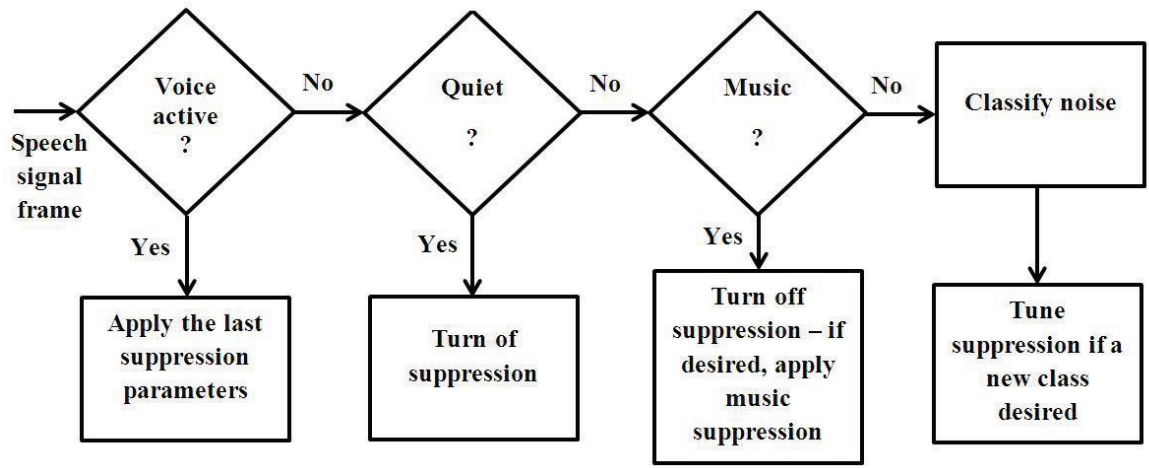

Figure 5.1. Addition of quiet and music detection capabilities.

*Adding quiet detection functionality to the VAD*

The function of detecting quiet segments is added into the VAD (Gopalakrishna, et al. 2012; Stadtschnitzer, Pham and Chien 2008) by changing the threshold used in the VAD for distinguishing speech from noise.

Among different VAD systems proposed in the literature (Stadtschnitzer, Pham and Chien 2008; Ramírez, et al. 2005; Nemer, Goubran and Mahmoud 2001), those utilizing wavelet packet transform is considered here as this transform is already computed as part of the CI speech processing pipeline. This choice ensures that the VAD component will not incur significant computational burden on the overall system. The approach used in (Stadtschnitzer, Pham and Chien 2008) is thus adopted here where speech and noise frames are distinguished based on Subband Power Difference (SPD) between lower and higher frequency bands. This difference is computed using the wavelet coefficients from the first level WPT coefficients of the input frame. This step is then followed by a signal power based weighting, a compression mapping and a first



order lowpass filtering to smooth out fluctuations. A speech or noise decision is made based on a thresholding procedure on the smoothed compressed subband power difference.

Let $\psi_{1,m}^0(n)$ and $\psi_{1,m}^1(n)$ denote the first level wavelet coefficients in the lower and higher frequency bands, respectively. Then, the SPD in each frame $m$ is computed as follows:

$$SPD(m) = \left| \sum_{n=1}^{N/2} (\psi_{1,m}^0(n))^2 - \sum_{n=1}^{N/2} (\psi_{1,m}^1(n))^2 \right| \tag{5.1}$$

where $N$ indicates the total number of samples in each analysis window. If $p_m(n)$ is the corresponding input speech signal power function, the SPD is weighted according to (5.2), followed by a compression according to (5.3) to obtain a smoothed compressed sub-band power difference in frame $m$ denoted by $Dc(m)$.

$$Dw(m) = SPD(m) \left[ 0.5 + (\frac{16}{\log(2)}) \log(1 + 2 \sum_{n=1}^{N} p_m(n)) \right] \tag{5.2}$$

$$Dc(m) = (1 - \exp(-2Dw(m))) \Big/ (1 + \exp(-2Dw(m))) \tag{5.3}$$

$Dc(m)$ values are stored in a buffer of size $B$, ordered in an ascending order in $Dcs$, then used according to (5.4) to set an adaptive threshold $Tv(m)$ when the condition in (5.5) is met. The threshold is updated as in (5.6) with $\alpha_v = 0.975$.

$$Tv(m) = Dcs(b) \tag{5.4}$$

$$Dcs(b) - Dcs(b-4) > 0.008, \quad \forall b = 4,...,B \tag{5.5}$$

$$Tv(m) = \alpha_v Tv(m-1) + (1 - \alpha_v) Tv(m) \tag{5.6}$$



If the smoothed compressed sub-band power difference $Dc(m)$ is less than $Tv(m)$ and more than a fraction of $Tv(m)$ by a coefficient $k_Q$ between 0 and 1, the frame is considered to be noise and if $Dc(m)$ is less than $Tv(m)$ by a factor $k_Q$, it is considered to be quiet, that is

$$\begin{cases} Dc(m) \geq Tv(m), & \text{Voice} \\ k_Q.Tv(m) \leq Dc(m) < Tv(m), & \text{Noise} \\ Dc(m) < k_Q.Tv(m), & \text{Quiet} \end{cases} \qquad (5.7)$$

For $k_Q = 0$, the VAD is reduced to the original VAD without the capability to distinguish quiet frames, as shown in Figure 5.2. For $k_Q = 1$, the VAD treats all noise frames as quiet, which is shown in Figure 5.3. Via experimentation, it was found that a value close to zero leads to a consistent detection of the quiet condition.

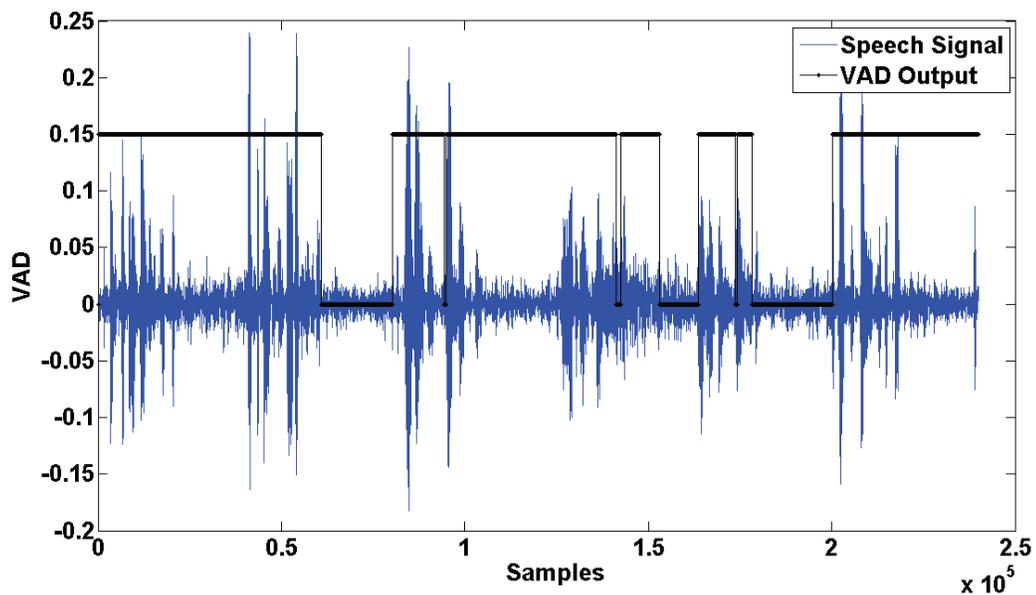

Figure 5.2. Modified Voice Activity Detector (VAD) output on a sample speech signal for threshold coefficient of zero (zero output indicates noise, positive indicates voice segments).



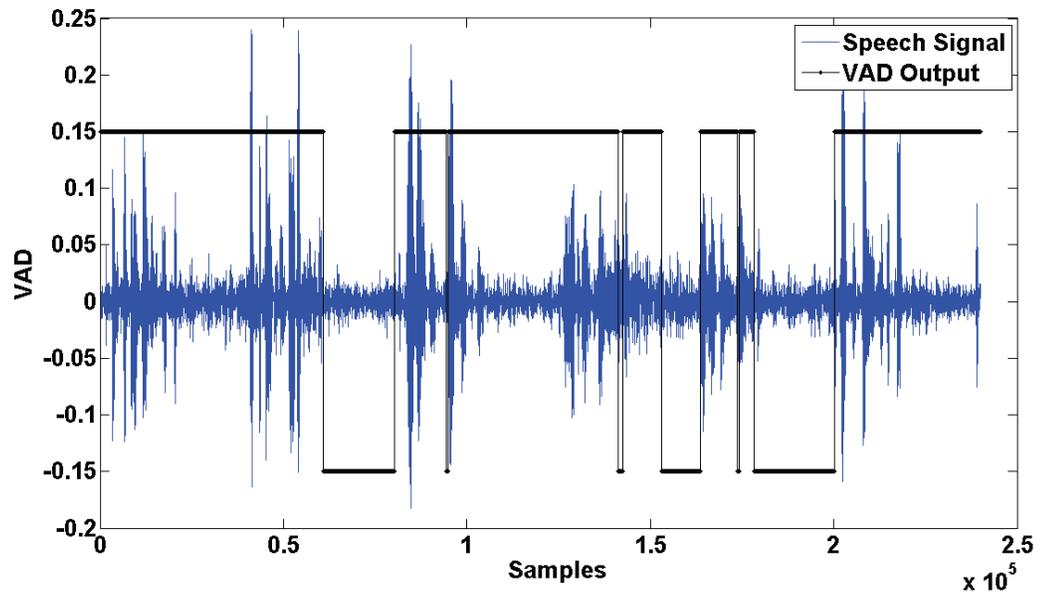

Figure 5.3. Modified Voice Activity Detector (VAD) output on a sample speech signal for threshold coefficient of unity (negative output indicates quiet, positive indicates voice segments).

### 5.3.2 Music detection

It is possible to use the same approach previously used to classify and categorize different types of noise for the purpose of distinguishing music from noise. This time, however, the training and testing of a GMM classifier is done for the purpose of music/noise classification rather than noise categorization. This is a two-class classification problem distinguishing music from noise such that any kind of noise is placed into the noise class and any kind of music is placed into the music class. The GMM classifier is therefore a simpler one than the noise categorization GMM classifier. Notice that high music/noise classification rates are deemed necessary for having an acceptable performance of the pipeline.



In (Gopalakrishna, et al. 2012), it is empirically shown that a 26-dimensional feature vector including a combination of MFCCs with their first derivatives provides a high average classification rate while not being computationally intensive. 40 overlapping triangular filters are used to map the 64-frequency bands magnitude spectrum of the wavelet packet transform to 40 bins in mel scale frequency. The lowest frequency considered is 133 Hz, and the first 13 consequent filters are spaced linearly with a bandwidth of 66.66 Hz, the remaining 27 filters are placed such that the bandwidths increase logarithmically with the highest frequency being 4000 Hz. A discrete cosine transform (DCT) is then applied to the logarithm of the magnitude spectrum in mel scale thus generating 13 MFCCs. If $MFCC(m, p)$ is the $p$-th MFCC coefficient in frame $m$, another 13 features from the first derivatives of the coefficients as computed in (5.8), constitute the 26-dimensional feature vector used for classification.

$$\Delta MFCC(m, p) = MFCC(m, p) - MFCC(m-1, p) \qquad (5.8)$$

Basically, a hierarchical approach is devised in this dissertation to add the music/noise detection capability to the pipeline. First, the VAD output is used to determine whether a current frame is quiet, noise or voice. If it is quiet, the suppression algorithm is turned off as one does not wish to introduce any distortion into speech when the noise level is so negligible that it gets detected as quiet. If it is detected as a voice-active frame, the latest settings of the suppression algorithm are kept. Only when the VAD outputs a noise frame, it is considered to be pure noise or containing music. In other words, the music/noise classification is done in order to decide the type of enhancement processing in the subsequent components.

# CHAPTER 6

## EXPERIMENTAL RESULTS AND DISCUSSION

This Chapter provides the experimental results corresponding to different bilateral extensions in the environment-adaptive pipelines of cochlear implants and the environment detection improvements as well as their discussions. Section 6.1 discusses some new measures to assess the environment detection improvements. It also provides an introduction to the objective speech quality prediction measures. The recorded noise data and speech and Head-Related Transfer Function databases used in the experiments are described in Section 6.2. Finally, the experimental results of different algorithms, pipelines and extensions that are developed in this dissertation are discussed in Section 6.3.

## 6.1    Performance evaluation

This Section covers the introduced quiet detection and suppression advantage measures. A review of speech quality prediction methods is also provided.

### 6.1.1    Existing speech quality and intelligibility measures

Subjective listening tests are the most accurate quality evaluation measures, but are time consuming, costly and require trained listeners (Hu and Loizou 2007; Loizou 2011) (at least need to train subjects on how to unbiasedly rate the enhanced speech, need specific significance statistical tests, need to address issues concerning the reliability of rater confidence assessments





usually referred to as intra- and inter-rater (Loizou 2006)). Many objective measures attempt to predict subjective quality as perceived by the human auditory system (Loizou 2011). Designing accurate subjective evaluation measures require involved psychoacoustical studies on the human auditory system (Loizou 2011). For example, it has been shown that the distance measures should not be symmetric and the measure associated with mapping should not be uniform over frequency bands.

Among the most commonly used objective evaluation methods (Quackenbush, Barnwell and Clements 1988), some are based on LPC coefficients such as the log-likelihood ratio (LLR), Itakura-Saito (IS) distance measure, cepstrum distance measure (CEP), cepstrum coefficients (Kitawaki, Nagabuchi and Itoh 1988), time domain measures such as segmental SNR (Hansen and Pellom 1998) and its frequency domain counterpart, the frequency weighted segmental SNR (Tribolet, et al. 1978), and perceptual evaluation of speech quality (PESQ), which is an ITU-T recommended standardized objective quality measure (ITU 2000).

Experiments in (Hu and Loizou 2008) reported the correlation of different objective evaluation measures with subjective test results reported in (Hu and Loizou 2006) for 13 different noise suppression algorithms. The evaluations were performed in terms of three different distortions perceived as signal distortion itself (SIG), background intrusiveness alone (BAK) and overall quality scores (OVRL). According to their correlation analyses, PESQ, LLR and the frequency weighted segmental SNR measures were found to be the most reliable ones. PESQ consistently showed the highest correlation with the subjective scores for all three ratings of SIG, BAK and OVRL (Hu and Loizou 2008). Although PESQ is viewed as the most reliable one, it is computationally demanding. It was also concluded that the time-domain segmental SNR measure



is not suitable for speech enhancement applications (Hu and Loizou 2008). It should be mentioned that the segmental SNR was first designed for quality assessment in speech coding applications (Quackenbush, Barnwell and Clements 1988). The study in (Hu and Loizou 2008) also suggests a modification of the PESQ measure which results in even higher correlation with subjective results.

Since using a single objective measure as the sole evaluation metric would not be capable of addressing all types of distortions from a practical standpoint, one could expect to achieve better subjective predictions by appropriately combining different measures. In the so called composite measures (Hu and Loizou 2008), linear regression analysis is usually used to find a more reliable measure by a linear combination of different single measures. Nonlinear combinations could result in even more correlated predictions (Hu and Loizou 2008).

### 6.1.2   Suppression advantage

In this Section, a new measure named Suppression Advantage is defined in order to quantify the noise suppression improvement of an entire pipeline due to noise classification. This measure provides a quantitative score for a joint performance of the noise detection and the noise suppression paths of any environment-adaptive speech enhancement pipeline.

*Definitions*

Let $\mathbf{P} = [P_{ij}]_{N \times N}$ be the confusion matrix associated with the above classifier, where $N$ is the total number of environment classes and let

$$P_{ij} \triangleq P(C_i \mid C_j) \tag{6.1}$$

be the probability that the classifier decides class $C_i$ while the true class is $C_j$ with



$$\sum_{i=1}^{N} P_{ij} = 1. \tag{6.2}$$

Also, let $\mathbf{Q} = [Q_{ij}]_{N \times N}$ be the quality matrix associated with the noise suppression component, where

$$Q_{ij} \triangleq Q(C_i \mid C_j) \tag{6.3}$$

denotes the quality measure achieved when using the suppression parameters associated with class $C_i$ while the true class is $C_j$.

Based on the above definitions, the expected quality for each class can be defined as follows

$$\overline{Q}_j \triangleq \sum_{i=1}^{N} P_{ij} Q_{ij}, \quad \forall j = 1, \dots N. \tag{6.4}$$

By writing $\mathbf{Q}$ and $\mathbf{P}$ as these matrices

$$\mathbf{Q} = [\mathbf{Q}_1, \dots, \mathbf{Q}_j, \dots, \mathbf{Q}_N], \tag{6.5}$$

$$\mathbf{P} = [\mathbf{P}_1, \dots, \mathbf{P}_j, \dots, \mathbf{P}_N], \tag{6.6}$$

(6.4) can be written as

$$\overline{Q}_j = \mathbf{P}_j^T \mathbf{Q}_j, \quad \forall j = 1, \dots N. \tag{6.7}$$

The overall expected quality of the pipeline can then be stated as

$$\overline{Q} = \sum_{j=1}^{N} P_0(C_j) \overline{Q}_j. \tag{6.8}$$

where $P_0(C_j)$ denotes the prior probability of class $C_j$.



*Fixed and Adaptive Expected Quality*

The expected values of different classes, $\overline{Q}_j$'s, depend on both the classifier and suppression components of the pipeline, thus $\overline{Q}$ evaluates the joint performance of the classifier and suppression components. Now, it is of interest to know how utilizing a noise classifier in the pipeline translates to a better suppression performance of the entire pipeline. To answer this question, a measure named Suppression Advantage (SA) is introduced that quantifies the amount of improvement in quality measure when using an environment-adaptive suppression pipeline. This measure allows one to quantify how the overall performance improves when the classification performance improves.

Let $\overline{Q}\{A\}$ be the expected quality associated with the adaptive suppression pipeline using a noise classifier with a confusion matrix of **P** as defined in (6.4), and $\overline{Q}\{F\}$ correspond to the fixed suppression using the same fixed suppression parameter set for all noise classes, then

$$\overline{Q}\{A\} = \sum_{j=1}^{N} P_0(C_j).\overline{Q}_j\{A\}, \tag{6.9}$$

$$\overline{Q}\{F\} = \sum_{j=1}^{N} P_0(C_j).\overline{Q}_j\{F\}. \tag{6.10}$$

For adaptive suppression, from (6.7) and (6.9),

$$\overline{Q}_j\{A\} = \mathbf{P}_j^T \mathbf{Q}_j\{A\}, \quad \forall j = 1,...N. \tag{6.11}$$

For fixed suppression,

$$Q_{ij}\{F\} = \text{constant} = Q_j\{F\}, \quad \begin{aligned} &\forall i = 1,...,N, \\ &\forall j = 1,...,N. \end{aligned} \tag{6.12}$$

Therefore, based on (6.2) and from (6.4),



$$\overline{Q}_j\{\text{F}\} = Q_j\{\text{F}\}, \quad \forall j = 1,...N. \tag{6.13}$$

It can be easily seen that $\overline{Q}_j\{\text{F}\}$ is independent of the confusion matrix, i.e. the expected quality is independent of the classifier performance.

*Suppression Advantage Measure*

To quantify the quality improvement, a base expected quality measure value is computed when there is no suppression, and then the SA measure is defined as the amount of increase in the quality measure for fixed or adaptive suppression pipelines.

Let $\overline{Q}_j\{\text{N}\}$ be this base quality measure value corresponding to the noise class $C_j$. This value is the one given by the quality measure $Q$ when no suppression is performed on speech signal.

Then, SA of an environment- adaptive or fixed pipeline with respect to the quality measure $Q$ can be stated as

$$SA^Q\{\text{A}\} \triangleq \overline{Q}\{\text{A}\} - \overline{Q}\{\text{N}\}, \tag{6.14}$$

$$SA^Q\{\text{F}\} \triangleq \overline{Q}\{\text{F}\} - \overline{Q}\{\text{N}\}. \tag{6.15}$$

Furthermore, it can be easily derived that the suppression advantage of a pipeline for each noise class is

$$SA_j^Q\{\text{A}\} = \overline{Q}_j\{\text{A}\} - \overline{Q}_j\{\text{N}\}, \quad \forall j = 1,...N, \tag{6.16}$$

$$SA_j^Q\{\text{F}\} = \overline{Q}_j\{\text{F}\} - \overline{Q}_j\{\text{N}\} = Q_j\{\text{F}\} - Q_j\{\text{N}\}, \quad \forall j = 1,...N. \tag{6.17}$$

### 6.1.3 Quiet detection assessment

A simple performance evaluation metric is defined here to quantify the accuracy of quiet detection performed by the modified VAD. Let



$$Q(m) = \begin{cases} 0, & \text{if } m \text{ not a quiet frame} \\ 1, & \text{if } m \text{ a quiet frame} \end{cases} \tag{6.18}$$

denote a function indicator for a given frame $m$ being an actual quiet frame. Similarly, let

$$\hat{Q}(m) = \begin{cases} 0, & \text{if } m \text{ not detected to be a quiet frame} \\ 1, & \text{if } m \text{ detected to be a quiet frame} \end{cases} \tag{6.19}$$

denote the corresponding function indicator estimated by the VAD. The VAD performance for

quiet detection is perfectly accurate when $\hat{Q}(m) = Q(m)$. However, it reflects zero accuracy when

$\hat{Q}(m) \neq Q(m)$ over the frames $m = 1, 2, ..., M$. Therefore, the quiet detection performance metric is

defined as the similarity between these two functions defined as follows

$$P_Q \triangleq \blacksquare \frac{1}{M} \sum_{m=1}^{M} \left| Q(m) - \hat{Q}(m) \right| \tag{6.20}$$

As a result, noting that $0 \leq P_Q \leq 1$, values close to unity indicate more accurate performance.

## 6.2     Real environment experiments and noise, speech and HRTF databases

In this Section, the real noise data recorded in different environments, clean speech and Head-

Related Transfer function databases used in the experiments are described.

### 6.2.1     Real environment noise recordings

Six commonly encountered noise types are considered in the experiments. Noise samples were

collected using the same BTE (Behind-The-Ear) microphone worn by Nucleus ESPrit implant

users. For each environment, a total of five sample files of one minute duration were collected.

In every recording, the integrated (average) sound pressure levels (SPLs) were also data-logged

for the run periods of one minute, almost exactly while the BTE recordings were on. The average



SPLs were 75.8 dBA for Street, 66.4 dBA for Car, 60.4 dBA for Restaurant, 67.8 dBA for Mall, 81.7 dBA for Bus, and 74.6 dBA for Train noise.

### 6.2.2 Speech data

Clean speech signals used in trainings for the single-processor gain parameters and comparisons were from IEEE Corpus database (IEEE Subcommittee 1969).

### 6.2.3 HRTF database

The CIPIC HRTF database (Algazi, et al. 2001) was utilized to generate the reference and non-reference signals associated with the clean, noise and noisy data. It was considered that both the clean and noise signals passed through a left and a right HRTF before being delivered to the CI speech enhancement pipeline. The subject number 3 of the CIPIC database is used. The elevation angle was assumed to be zero, but different azimuth angles ranging from -80 to 80 degrees with this discretization [-80, -65, -55, -45:5:45, 55, 65, 80] were considered for a total of 25 different Head-Related Impulse Responses (HRIRs). Half of the data covering -80 to 0 degrees (a total of 13 HRIRs) were used; for the other half the only difference was switching the reference signal with the non-reference signal.

### 6.3 Results and discussion

In this Section, the results of the experiments using the developed noise suppression and noise detection algorithms for the bilateral environment-adaptive pipelines of cochlear implants (CIs) are also presented and discussed.



### 6.3.1 Single-processor bilateral speech processing pipeline

The experimental setup for the bilateral extension developed in Chapter 2 is described and the results are discussed in this Section.

*Experimental setup*

Speech quality and timing performance of the discussed extension were assessed using six commonly encountered noise types, IEEE sentences and CIPIC HRTF database as explained in Section 6.2. To simulate bilateral hearing conditions, speech sentences were convolved with a patient-specific HRTF. Training of the suppression functions along with the HRTF gain parameters were performed using the first 50 clean speech sentences from the IEEE Corpus database. All suppression and HRTF gain parameters associated with each environment were trained by adding the recorded noise samples to the IEEE sentences as clean speech signals. The resulting noisy files were then used to generate the required training set.

The estimation of the HRTF gains were parameterized in seven different ITD values, distributed uniformly. Each $\hat{\tau}$ estimate was assigned to its closest HRTF parameter point. These impulse responses were downsampled to the clean signal sampling frequency. They were then convolved with the clean, noise and noisy signals to simulate pairs of reference and non-reference signals. Although only a portion of the entire spherical space around the head was examined here, other elevation or azimuth angles can be simply evaluated in a similar way to generate corresponding parameters. For higher resolution elevation and azimuth angles for which the HRIRs were not available, a simple linear interpolation technique was used to estimate the enhancement output in the missing points based on their neighboring HRIRs.



The performance of the gains resulted from the joint optimization approach is compared against using the direct one-channel gain optimization applied independently on each of the reference and the non-reference signals. The weighted Euclidean distortion measure was used for a fair comparison. A total of three out of five noise files for each environment and the first 50 IEEE sentences were used to generate the training data. Two different gain tables were obtained using the reference data for the left ear and using the non-reference data for the right ear. The combination of the same reference and non-reference data was used in the two-channel speech processing pipeline involving a single-processor to obtain a suppression gain and an HRTF function for 13 different azimuth angles of the six different noise types. It was always assumed that the left ear was ipsilateral to the sound source, hence receiving the reference signal.

Noting that the delay estimation itself could produce errors in each frame, a median filter across the past 20 delay estimations was used to filter out outlier estimates. This helps to prevent clicks in the outputs when the direction between the ears and the source changes. By using the median filter, a smoother transition for the change of direction over time was obtained.

Although subjective listening tests provide accurate quality evaluation measures, they are time consuming, costly and require trained listeners as discussed in Section 6.1.1. Fortunately, there are many objective evaluation methods (Quackenbush, Barnwell and Clements 1988) attempting to predict the subjective quality of processed speech. Perceptual Evaluation of Speech Quality (PESQ), an ITU-T recommended standardized objective quality measure (ITU 2000), is a widely used measure for this purpose. It was utilized here for evaluation purposes.

It is worth noting that the developed joint enhancement approach performs the noise estimation and a priori SNR estimation only on the reference signal, but the non-reference counterpart is



reconstructed based on the direction or ITDs, and thus this is a more efficient computation than processing the two signals.

<p align="center">*Results and discussion*</p>

Figure 6.1 (a-f) show the comparison of the PESQ measure evaluating the optimized joint gain parameters using single-processor bilateral processing against those involving one-channel enhancement gains using double-processor bilateral processing of the reference and non-reference signals independently for the six different noise types across 13 azimuth angles ranging from -80 to 0 degrees. The PESQ scores on the non-processed noisy signals are shown to serve as the baseline. The results shown denote averages on the second half of the 50 IEEE sentences which had not been considered in the training phase. The average gain in PESQ in each environment over all the 13 angles are also shown in these figures. Also, the other two noise files of each environment not used in the training were used to generate noisy signals. It can be observed that for the reference signal which was obtained directly by applying the suppression gain table, there was not a significant performance loss when using our joint optimization method. However, as indicated by the average gains in PESQ, a higher quality was achieved for the non-reference signal which was obtained by applying the HRTF estimates onto the reference counterpart according to the binaural reconstruction based on the time delay estimation. From Figure 6.1, it can be seen that not only one gets the binaural stimulation capability, but also some correction of errors given by the single suppression gain table. In other words, one obtains the non-reference processed signal with a slightly higher PESQ scores than the single optimization of the suppression gain.



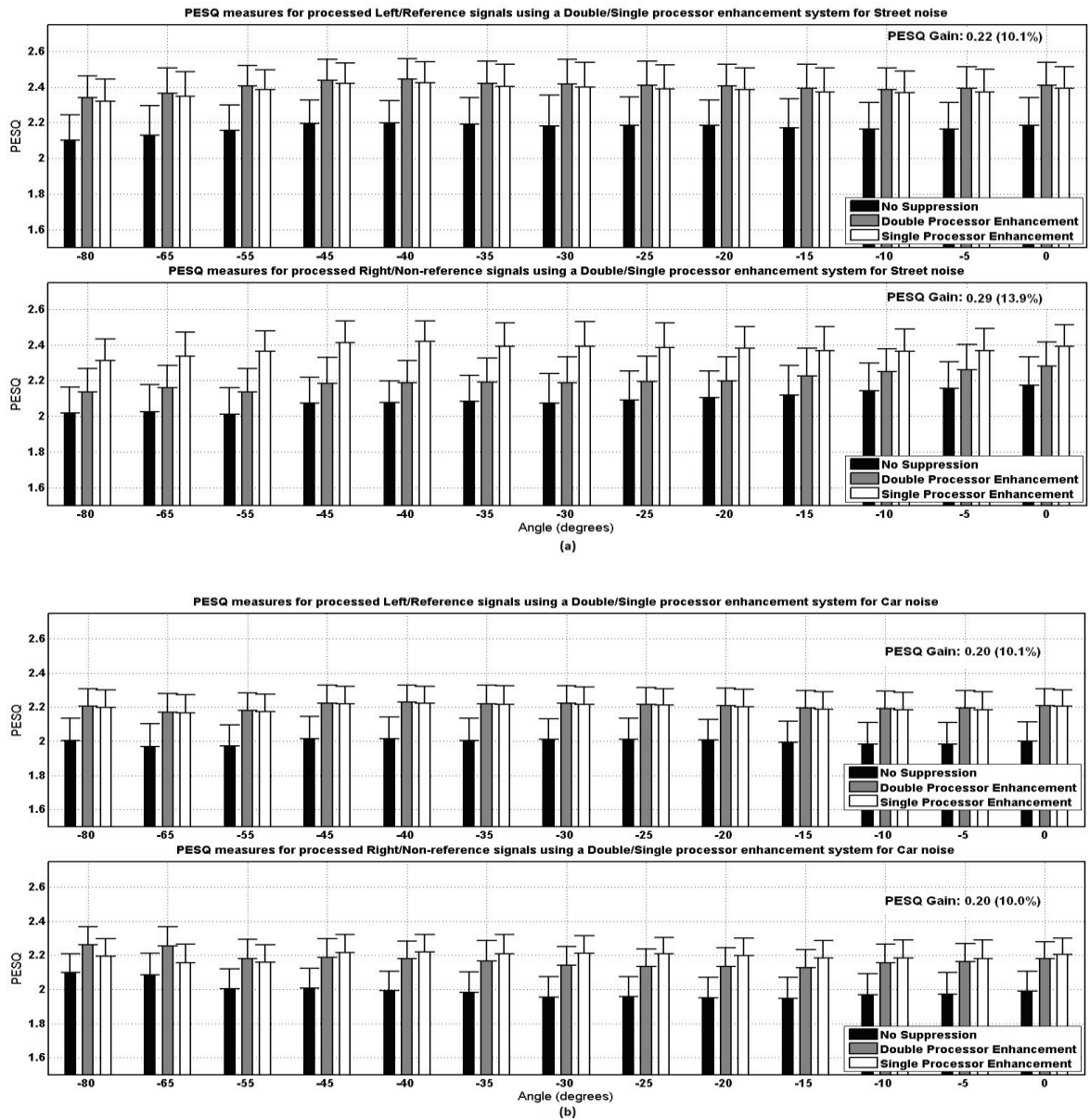

Figure 6.1. Quality assessments are performed based on Perceptual Evaluation of Speech Quality (PESQ) scores, comparison with no-suppression scores as baseline in six noisy environments of (a) Street, (b) Car, (c) Restaurant, (d) Mall, (e) Bus and (f) Train noise. The double-processor counterpart processes the left and right signals independently in parallel, each using one of the two available processors. The suppression gain tables used for the left and right signals were optimized by training over their associated left or right collected data along each direction. The same datasets were used for training the developed single-processor pipeline and Head-Related Transfer Function (HRTF) gain tables. The left signal was considered to be the reference input in this set of experiments.



Figure 6.1. Continued.

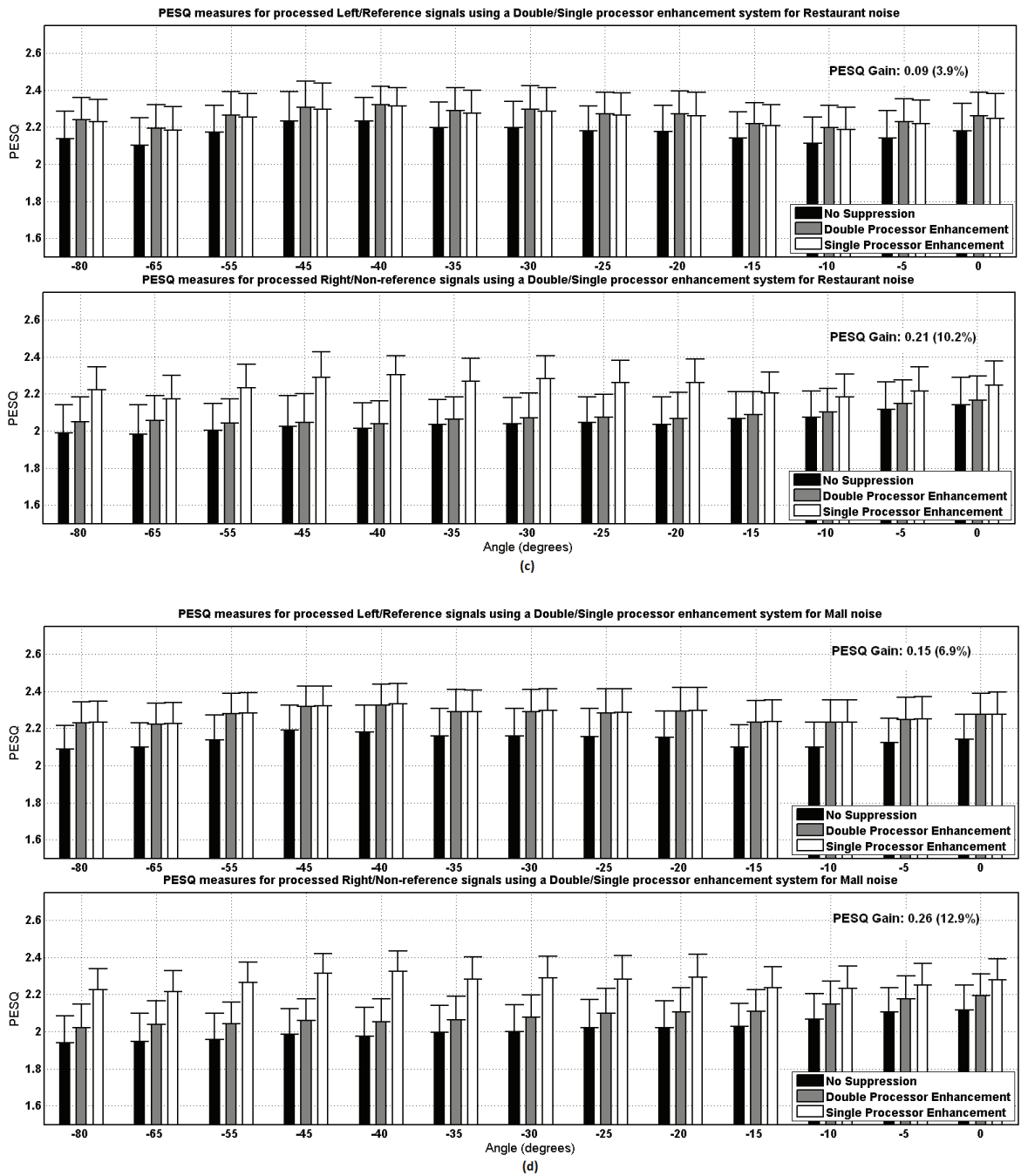



Figure 6.1. Continued.

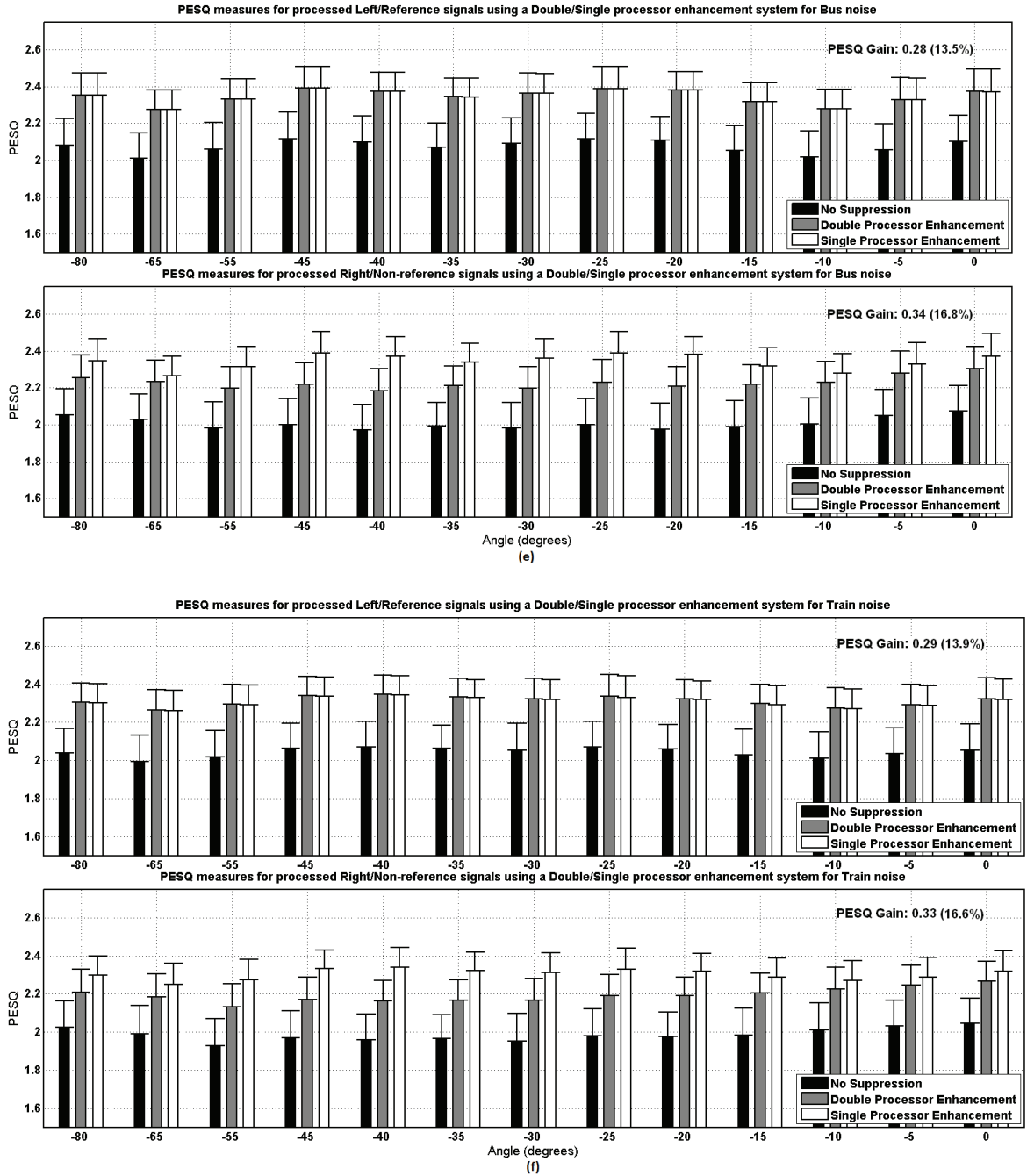

(e)

(f)



A statistical significance analysis using the standard t-test was carried out revealing that in all 78 cases involving six noise types and 13 directions, it always failed to reject the null hypothesis between double-processor and single-processor PESQ scores on the reference signal. The averages of the p-values over different directions were 0.44, 0.76, 0.89, 0.96, 0.67 and 0.83 for the noisy environments of Street, Car, Restaurant, Mall, Bus and Train, respectively, indicating non-significant differences. On the other hand, the same statistical test on the associated non-reference PESQ scores showed significant improvement of single-processor results over double-processor ones in 70 out of 78 cases at the 95% confidence level and in 65 cases at the 99% confidence level.

As shown in Table 6.1, the CPU processing time was significantly decreased by using the single-processor pipeline (the timing measurements were obtained on a PC platform with a CPU of 2.66 GHz clock rate. Also, note that the numbers denote average timings for processing 50 IEEE sentences which are approximately 2-3 seconds long). It can be noticed that the independent processing via using two processors running in parallel is only 11% faster than using only a single processor. Furthermore, the developed single-processor approach is about 44% faster than using a single processor with a sequential processing of the two signals. Thus, it is more suitable for deployment than a sequential processing of both the left and right signals for bilateral stimulation. It is also worth emphasizing that in the developed approach there would be no need to be concerned about any synchronization to coordinate the operation of the two signal processing paths.



Table 6.1. Average timing outcome over all noise types and all azimuth angles (in seconds on 50 speech files with length of 2-3 seconds) by a single processor (proposed) compared to sequential independent processing of the left and right signals using a single processor and a double-processor system (two separate processors) without any synchronization.

| Hardware System Architecture / Direction | Left | Right |
|---|---|---|
| Double-Processor | 0.2741 | 0.2735 |
| Single-Processor (Sequential Processing) | 0.5476 | |
| Single-Processor (Proposed) | <u>0.3081</u> | |

Six commonly encountered noise types have been considered and the gain tables were for each environment independently. When the ambient noise type is not one of these six trained environments, the classifier assigns it to the closest class among the trained noise types and the associated suppression gain table is loaded to the enhancement component. To assess the performance when a noise type other than those considered for training is encountered, an unknown noise (Flight noise) was added to the clean speech files, but the trained gain tables were used for the enhancement. Figure 6.2 shows the average PESQ measures for the left and right processed signals when each of the six trained gain tables was used for the enhancement. It can be observed that the measures were still improved but not as much as when the specific gain tables were used for each specific noise type. This experiment demonstrates how unseen noise types are dealt with. It should be emphasized that the measures in Figure 6.2 correspond to the averaged outcome over all the trained gain tables. However, in general, better results are obtained as the gain for the noise type with the highest similarity to the unknown noise is loaded into the enhancement function.



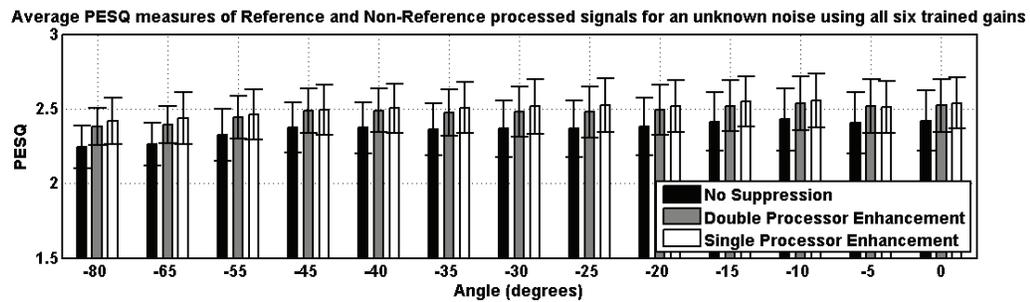

Figure 6.2. Perceptual Evaluation of Speech Quality (PESQ) comparisons when an unknown noise is encountered. The scores are the average of the right and left quality measures when each of the other six trained gain tables is used for the enhancement.

In summary, the introduced pipeline is computationally more efficient than independently processing two signals for bilateral stimulation while it does not cause any statistically significant performance loss in terms of speech quality. Meanwhile, it is still environment-adaptive and computationally efficient. These characteristics make this pipeline a suitable choice for deployment in bilateral CIs. It is worth mentioning that the framework is data-driven and is not based on a closed form analytical solution. If nonlinear optimization techniques are used, various distortion measures can be considered. Also, it should be noted that putting both suppression and HRTF gain functions in the data-driven optimization enables some modeling and estimation errors in the suppression gain to be corrected by the HRTF parameters. Finally, it is to be emphasized that a simple extension of the previous one-channel pipeline to include directionality has not been considered, rather a computationally tractable framework is reached by not considering different suppression gains for different directions, but by training the HRTF parameters directly in Chapter 2. This makes the obtained optimized parameters reusable for other speech processing tasks.




*Summary*

Because the optimization of suppression and binaural reconstruction gain parameters of the framework developed in Chapter 2 is done in a data-driven manner, it can be tuned to different noise types with different characteristics. The PESQ quality assessment measure in six different commonly encountered noisy environments showed that the performance loss is statistically not significant compared to the independent processing of the left and right signals. The developed pipeline has been demonstrated to be efficient in computations and storage requirements.


### 6.3.2 Generalizations to non-Euclidean distortion criteria

The generalized bilateral pipeline developed in Chapter 3 provides optimization solutions to find suppression and HRTF gain parameters based on non-Euclidean distortion criteria. In this Section, experiments are reported to evaluate this generalization based on different distortion measures.

*Experimental setup*

Having extracted solutions in (3.8-3.9) for the WE, in (3.13-3.14) for the LE, and in (3.20-3.21) for the WC distortions, any gradient-based non-linear optimization method can be used to train the gain parameters. Here, the simple steepest descent together with a momentum-based learning rate adaptation (used a momentum multiplier of 0.9) was used. Learning rates were considered to be 0.5, 1e-6 and 5e-7 for WE, LE and WC, respectively. All the common settings were chosen the same as the ones in (Mirzahasanloo, et al. 2013) and Section 6.3.1 for comparison purposes. The IEEE sentences were used as clean speech signals and the CIPIC HRTF dataset was used to generate the HRTF-convolved reference and non-reference noise, noisy and clean speech training and testing data. The CIPIC data for 13 different azimuth angles at 0° of elevation was



used for training and testing. Noise data were recorded using the BTE microphones in real environments with the PDA research platform in six commonly encountered noise environments of Street, Car, Restaurant, Mall, Bus and Train, and then were added to the clean speech signals at 5 dB SNR.

*Results and discussion*

Table 6.2 and Table 6.3 show the PESQ scores for each noise environment averaged over reference and non-reference outcomes ($\beta = 1$) and over 13 different angles. In each test case, 50 IEEE speech files (not seen during training) were used (total of 650 test samples for each environment).

Segmental SNR improvements are also presented here to show how each method reduced noise levels. It can be seen that although WC provided the highest SNR improvements, it did not reach the highest quality scores except in Restaurant and Train environments. These differences were statistically significant at 99% confidence level. This implied that WC reduced noise more than the other methods but also caused removal of parts of speech, thus introducing distortions and causing speech quality loss. WE and LE did not result in significant SNR+ or PESQ score differences, but both provided higher PESQ scores in Mall and higher SNR+ in Street, Mall and Train than the direct estimation method in (Mirzahasanloo, et al. 2013) (Dir) at 95% confidence level and in all the other environments at 99% confidence level. Using a noise environment recognition approach such as the ones in (Mirzahasanloo, et al. 2013; Mirzahasanloo, et al. 2012), the best performing gain for each environment was loaded to the pipeline suppression component.



Table 6.2. Segmental Signal to Noise Ratio (SNR) improvements and Perceptual Evaluation of Speech Quality (PESQ) scores for different methods of direct quasi-static gain estimation (Dir), gradient-based training based on Weighted-Euclidean (WE), Log-Euclidean (LE) and Weighted-Cosh (WC) distortion measures in Street, Car and Restaurant environments. Corresponding values for no suppression (N/S) are also shown for comparison.

| Noise Class | | *Segmental SNR+* | *PESQ* |
|---|---|---|---|
| | N/S | 0 | 2.13 (±0.15) |
| | Dir | **1.33** (±0.68) | 2.38 (±0.13) |
| *Street* | WE | **1.50** (±0.69) | **2.40** (±0.13) |
| | LE | **1.55** (±0.69) | **2.40** (±0.13) |
| | WC | **2.83** (±1.13) | 2.38 (±0.17) |
| | N/S | 0 | 1.99 (±0.12) |
| | Dir | 1.36 (±0.39) | 2.20 (±0.10) |
| *Car* | WE | **1.54** (±0.38) | **2.22** (±0.10) |
| | LE | **1.54** (±0.40) | **2.22** (±0.10) |
| | WC | **2.69** (±0.65) | 2.12 (±0.13) |
| | N/S | 0 | 2.08 (±0.14) |
| | Dir | 0.92 (±0.41) | 2.15 (±0.12) |
| *Restaurant* | WE | **1.12** (±0.42) | **2.18** (±0.12) |
| | LE | **1.10** (±0.42) | **2.18** (±0.12) |
| | WC | **2.84** (±0.78) | **2.23** (±0.14) |



Table 6.3. Segmental Signal to Noise Ratio (SNR) improvements and Perceptual Evaluation of Speech Quality (PESQ) scores for different methods of direct quasi-static gain estimation (Dir), gradient-based training based on Weighted-Euclidean (WE), Log-Euclidean (LE) and Weighted-Cosh (WC) distortion measures in Mall, Bus and Train environments. Corresponding values for no suppression (N/S) are also shown for comparison.

| Noise Class | | *Segmental SNR+* | *PESQ* |
|---|---|---|---|
| | N/S | 0 | 2.07 (±0.14) |
| | Dir | **1.58** (±0.43) | **2.27** (±0.12) |
| *Mall* | WE | **1.72** (±0.44) | **2.29** (±0.12) |
| | LE | **1.77** (±0.44) | **2.29** (±0.12) |
| | WC | **3.02** (±0.76) | 2.14 (±0.14) |
| | N/S | 0 | 2.04 (±0.14) |
| | Dir | 1.66 (±0.43) | 2.34 (±0.12) |
| *Bus* | WE | **1.84** (±0.42) | **2.36** (±0.11) |
| | LE | **1.84** (±0.44) | **2.36** (±0.11) |
| | WC | **4.14** (±0.72) | 2.31 (±0.17) |
| | N/S | 0 | 2.01 (±0.13) |
| | Dir | **1.79** (±0.47) | 2.31 (±0.11) |
| *Train* | WE | **1.94** (±0.45) | **2.33** (±0.11) |
| | LE | **1.97** (±0.48) | **2.32** (±0.10) |
| | WC | **4.38** (±0.60) | **2.34** (±0.13) |



### 6.3.3   Unified optimization framework

Additional experiments are also reported in this Section using the unified data-driven

optimization framework developed in Chapter 4 for the single-processor bilateral speech

processing pipeline covered in Chapter 2.

*Experimental setup*

The developed binaural speech processing framework in Chapter 4 provides a suitable bilateral

speech framework for environment-adaptive pipelines in unilateral and bilateral CI applications.

The suppression component and the HRTF parameters can be trained for different noise types

and used in the CI speech processing pipeline in different noisy environments that CI users

encounter in their daily lives. Not only the parameters can be tuned in different noisy

environments, the solutions obtained for different distortion criteria can be used in different

noisy environments where different criteria guide the speech processing enhancement. These

solutions also make it possible to study different speech distortion measures for speech quality

and intelligibility improvements in different conditions, for instance different noise

environments, different speech recognition or audio and music quality improvements as well as

different unilateral and bilateral user settings.

Speech quality improvements using the three developed bilateral solutions were evaluated based

on the three most commonly encountered distortion criteria in six most commonly encountered

noise environments. IEEE speech sentences were used as the clean speech dataset and to

generate noisy speech signals for training. Noise data were recorded using Nucleus ESPrit

Behind-The-Ear (BTE) microphones in the same way as worn by CI patients and using the FDA-

approved PDA research platform  (Mirzahasanloo, et al. 2012; Ali, et al. 2013) to data-log them



for training. All noise data were recorded in actual noise environments without any further processing, in order to best simulate realistic conditions under which the BTE microphone captures sound, noise and speech. Bilateral data were generated using the CIPIC HRTF database as described in Section 6.2.3 containing high resolution HRIR data. A training dataset was created by adding the actual recorded noise to the IEEE speech files and convolving the resulting noisy speech signals with the CIPIC HRIR filters. The HRIR data corresponding to the elevation angle of 0 degrees in 13 different azimuth angles were considered. Six different sets of suppression and HRTF gain parameters for six different noise environments were obtained after training the bilateral speech processing system using each corresponding noise data. Optimized gain parameters were obtained using each of the three distortion criteria of WE, LE, WC in each environment, therefore a total of 18 bilateral test cases were examined. 20 IEEE sentences were used for training while 50 were used for testing with no overlap with the training sentences. Similarly, 60% of the recorded noise data were used for training in each case and the rest were used for testing.

PESQ was used to evaluate the perceptually driven speech quality, and compared the noise reduction performance of the developed solutions with the previously developed solutions in (Gopalakrishna, et al. 2012; Erkelens, Jensen and Heusdens 2007) and the algorithms discussed in Chapter 2 and Chapter 3.

*Results and discussion*

Table 6.4 shows the comparison of PESQ between the developed bilateral pipeline and the previously developed solutions.



Table 6.4. Perceptual Evaluation of Speech Quality (PESQ) scores of Time Difference Of Arrival (TDOA) based direct (Dir) and gradient-based binaural reconstruction gains in single-processor bilateral speech processing based on Weighted-Euclidean (WE), Weighted-Cosh (WC) and Log-Euclidean distortion criteria compared with the developed methods of Interaural Phase Difference (IPD) based reconstruction in six most commonly encountered noise environments. Scores corresponding to No Suppression (N/S) on noisy speech are also provided for reference.

| PESQ (Binaural Reconstruction / Noise Type) | | *Street* | *Car* | *Restaurant* | *Mall* | *Bus* | *Train* |
|---|---|---|---|---|---|---|---|
| *N/S* | | 2.13 (±0.15) | 1.99 (±0.12) | 2.08 (±0.14) | 2.07 (±0.14) | 2.04 (±0.14) | 2.01 (±0.13) |
| | *Dir* | 2.38 (±0.13) | 2.20 (±0.10) | 2.15 (±0.12) | 2.27 (±0.12) | 2.34 (±0.12) | 2.31 (±0.11) |
| *Developed TDOA-binaural* | *WE* | 2.40 (±0.13) | 2.22 (±0.10) | 2.18 (±0.12) | 2.29 (±0.12) | 2.36 (±0.11) | 2.33 (±0.11) |
| | *WC* | 2.40 (±0.13) | 2.22 (±0.10) | 2.18 (±0.12) | 2.29 (±0.12) | 2.36 (±0.11) | 2.32 (±0.10) |
| | *LE* | 2.38 (±0.17) | 2.12 (±0.13) | 2.23 (±0.14) | 2.14 (±0.14) | 2.31 (±0.17) | 2.34 (±0.13) |
| *Developed IPD-binaural* | *WE* | 2.41 (±0.04) | 2.26 (±0.02) | 2.26 (±0.04) | 2.35 (±0.04) | 2.35 (±0.04) | 2.33 (±0.03) |
| | *WC* | 2.40 (±0.03) | 2.25 (±0.02) | 2.29 (±0.05) | 2.37 (±0.04) | 2.35 (±0.04) | 2.39 (±0.02) |
| | *LE* | 2.42 (±0.04) | 2.16 (±0.02) | 2.25 (±0.03) | 2.28 (±0.04) | 2.37 (±0.04) | 2.36 (±0.03) |



The results exhibit the outcomes corresponding to the bark scale-based sub-banding of the spectral domain used in the HRTF modeling via the IPD estimation and suppression gain estimation. The approach in (Mirzahasanloo, et al. 2013) and (Mirzahasanloo and Kehtarnavaz, 2013a) deployed a generalized cross-correlation based TDOA estimation and assumed an HRTF model using a function of different levels of time delays. Scores for the non-processed noisy speech files are also provided in this table for reference.

As can be seen from this table, the speech quality using the IPD-based binaural reconstruction methods improved in a consistent manner in different noisy environments and based on different distortion measures. The improvements were due to having higher resolution directional modeling in the developed IPD-based framework. As mentioned in Chapter 4, having all the processing in the spectral domain allowed avoiding the error-prone time domain delay estimations thus making the approach more robust and consistent. This is evident in the PESQ results as the standard deviation of the score given by the IPD-based methods is seen much lower than those provided by the TDOA-based ones. Also, note that the quality results are compared using the gain parameters when all the trainings were stopped at the same number of iteration. This was an early stop for the generalized methods as they involved a larger number of parameters. Naturally, further quality improvements may be achieved with longer training of the gain parameters, when the number of iterations of the offline training is not of concern.

The developed framework reported in Chapter 4 performs a spectral domain modeling and estimation of suppression and HRTF gain parameters, leading to a larger number of parameters to train. On the other hand, with some extra memory requirements, it provides significant computational and processing time advantages as no time-domain processing is required to



perform the binaural processing when using a single processor. This is a critical feature in implementation of CI speech processing pipelines as it allows generating high quality binaural signals with minimal added computations due to the second speech signal. The developed bilateral speech processing methods run using only a single processor requiring 0.5 processor per channel compared to 1 processor per channel when double-processor bilateral processing is used (Mirzahasanloo, et al. 2013). This provides 50% better hardware efficiency. As discussed in Chapter 2, this feature solves the synchronization problem as well. Applying the gain tables in the pipeline takes an average processing time of 0.40 seconds for IEEE speech sentences on a processor with 2.66 GHz clock rate. Compared to an average processing time of 0.54 seconds for sequential processing of bilateral signals or independent processing using two processors (Mirzahasanloo, et al. 2013), the developed enhancement is 25% faster. Also, with a word length of 16 bits used to store suppression and gain parameters, the introduced framework requires only about 8.7 KB of memory. Compared to the required 16.4 KB memory when sequential or double-processor independent processing is used (Mirzahasanloo, et al. 2013), it is about 47% more memory efficient. Hardware, computation and memory efficiency as well as no need for signal synchronization of bilateral stimulation makes the introduced framework a suitable solution for deployment in bilateral CIs.

A major advantage of the introduced unified data-driven model is that it enables not only having environment-specific suppression gains, but also user-specific and individualized HRTF models for binaural cue modeling whose parameters can be fitted to each user. The training data can be collected and data-logged by having users wear dual microphones in their daily experiences and use those logged data to fine-tune the gain parameters.



Currently used CI processors provide non-robust binaural cue estimations due to a lack of proper right and left signal synchronization. Just noticeable difference in ITD varies in different subjects and can be as low as 10 to 30 us depending on direction and the type of stimulus (van Hoesel 2007; Senn, et al. 2005). Achieving this resolution in time difference can be very challenging in the current devices that do not synchronize the left and right cochlea. The introduced approach enhances the binaural sensitivity without need for synchronization in the hardware device. With this generalized framework, by applying the binaural reconstruction gains in different decomposition bands after generating the stimulation pulses, it is now possible to greatly enhance the process of providing synchronized stimulation signals.


*Summary*

The generalized suppression and HRTF gain optimization framework developed in Chapter 4 for single-processor speech processing in bilateral cochlear implants using a single processor, allows the modeling of head-related transfer functions in the spectral domain and thus enabling the incorporation of hearing perception principles. Optimized solutions for two families of amplitude and loudness-weighted distortion measures or criteria have been derived and tested on six most commonly encountered noise environments. In addition to gaining speech quality improvements in a consistent way, the developed pipeline does not require any synchronization of the left and right signals. Furthermore, the pipeline is shown to computational and memory efficient. It has been shown that the previously developed unilateral and bilateral solutions covered in Chapter 1, Chapter 2, and Chapter 3 are special cases of this generalization. The data-driven nature of the solutions provides a highly customizable pipeline that can be optimized in different noisy




environments and individualized anthropometric measurements benefiting cochlear implant users.

### 6.3.4 Environment detection improvements

In this Section, the results of the dual-microphone noise classification in Section 5.2 and the addition of quiet and music detection capabilities in Section 5.3 are presented.

*Dual microphone classification results and discussion*

Noise data recorded by the BTE microphone worn by Nucleus ESPrit cochlear implant users sampled at a rate of 22050 Hz in four commonly encountered noise environments of Street, Car, Restaurant and Mall were used. These data were recorded in real noise environments using the FDA-approved PDA research platform for CI studies (Ali, et al. 2013; Mirzahasanloo, et al. 2012). In all the classification tests, 50% of the data were used for training and 50% for testing with no overlap between the training and testing data sets. The CIPIC HRTF database was also used to generate the left and right microphone signals as explained in Section 6.3.1. For each microphone signal, a 26-dimensional feature vector consisting of 13 Mel-Frequency Cepstrum Coefficients (MFCC) and 13 ΔMFCC features were used. For enhancement evaluations, the collected real noise data were used to generate noisy signals of the IEEE speech sentences. Table 6.5 compares the Correct Classification Rates (CCRs) using our dual-microphone classification and the feature-level approach with that of the previously developed single-microphone classification in (Gopalakrishna, et al. 2012). Using the dual-microphone classification, CCR improved by about 9.4%. Although using majority voting over a number of past classification decisions improved the classification performance considerably, time delays were introduced as a result of considering past decisions. The dual microphone approach allowed



lowering the number of past decisions leading to less time delays compared to the single microphone approach. As shown in Table 6.5, when using 10 frames for majority voting, 7% classification improvement was achieved while getting 50% less time delay. Note that this improvement became less pronounced as more frames or a longer history of past decisions was considered for majority voting at the expense of more time delay which ultimately limited the real-time operation of the entire pipeline.

Table 6.6 provides the feature extraction and classification processing times for 11.6ms speech frames on both the FDA-approved PDA platform with a 624 MHz clock rate as well as the PC platform with a 3.0 GHz clock rate while using the majority voting over past 20 frames. As can be seen from this table, the extra computation time due to the dual-microphone classification did not limit the real-time operation of the entire pipeline, i.e. the processing time stayed less than the frame length of 11.6ms (256 samples at 22050Hz sampling rate).

The dual-microphone approach also led to a better suppression performance of the environment-adaptive pipeline for all the noise classes as shown in Figure 6.3.

Table 6.5. Correct classification rates of dual-microphone classification compared to single-microphone classification for different number of past decisions or frames in majority voting.

| *Correct Classification Rate (%)* | **Without majority voting** | **With majority voting over last 10 decisions** | **With majority voting over last 20 decisions** |
|---|---|---|---|
| **Single-mic** | 74.3 | 81.6 | 91.5 |
| **Dual-mic** | 81.3 | 87.7 | 92.1 |



Table 6.6. Average timing profile of the entire pipeline for 11.6 ms frames (in ms).

| Platform | Total Time | A | B | C | D | E |
|:---:|:---:|:---:|:---:|:---:|:---:|:---:|
| PDA (single-mic) | <u>8.52</u> | 2.41 | *1.34* | 2.03 | *0.91* | 1.83 |
| PDA (dual-mic) | <u>10.39</u> | 2.41 | *2.62* | 2.03 | *1.80* | 1.83 |
| PC | <u>0.89</u> | 0.41 | 0.21 | 0.14 | 0.07 | 0.06 |

**A**: *FFT computation and suppression*; **B**: *Speech decomposition*; **C**: *VAD decision*;

**D**: *Feature extraction and classification*; **E**: *Channel envelope computation*.

The same rule for all the classes was used for fixed noise suppression and the ideal system was assumed to have a perfect classification accuracy. This figure shows the Suppression Advantage (SA) values defined in Section 6.1.2 with respect to PESQ. One can see that the dual-microphone approach provided better SA over the single microphone approach when using the environment-adaptive pipeline and also when using the fixed pipeline.

In summary, when using a dual-microphone, it was shown in Section 5.2 that the feature-level combination approach was more suitable for actual deployment than the decision-level combination approach due to its computational and memory efficiencies. It was also shown in this Section that the classification accuracy was improved as a result of using a dual-microphone compared to using a single microphone. A new measure named Suppression Advantage was also introduced in Section 6.1.2 to evaluate fixed and adaptive suppression pipelines of cochlear implants and it was shown that the dual-microphone classification provided better suppression advantage.



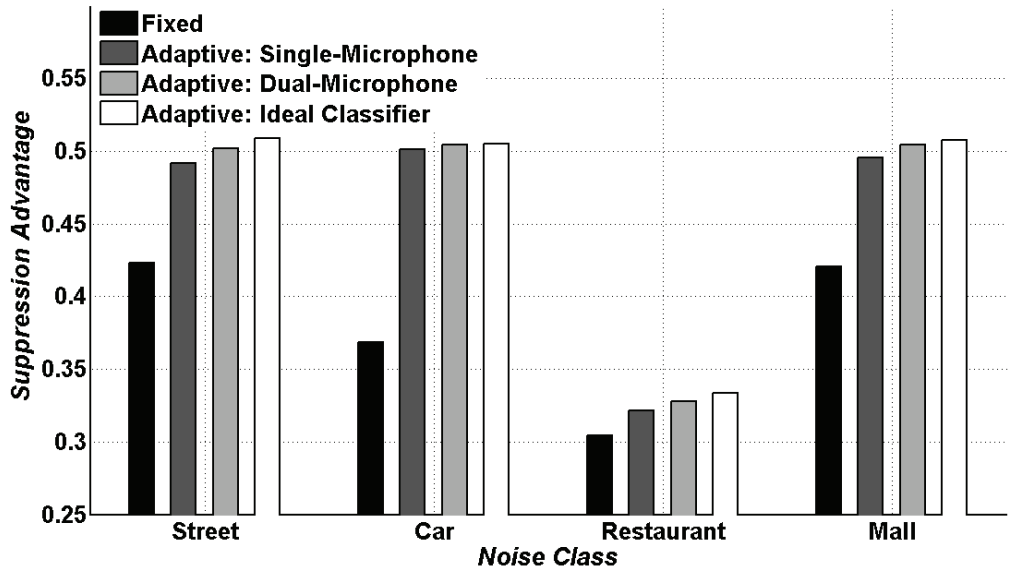

Figure 6.3. Suppression Advantage (SA) values with respect to the Perceptual Evaluation of Speech Quality (PESQ), for a fixed noise suppression using a log-MMSE (log-Minimum Mean Squared Error) estimator, environment adaptive suppression using single microphone classification, introduced approach using dual-microphone classification, and ideal classification for environment-adaptive suppression.

*Quiet and music detection results and discussion*

The modified VAD defined in (5.7) is evaluated by examining sample speech data files and the metric defined in (6.20). Figure 6.4 shows the performance of the system on a speech signal containing no–noise and voice segments.

Note that there was a small delay in detecting quiet segments that came immediately after noise or voice segments. This was due to not considering them quiet unless they had been detected as quiet for a number of times to ensure a robust performance. This was also the case when switching to a noise segment after a number of voice or quiet detections. This is in fact desired as one does not wish to label low energy parts of speech as quiet or noise.



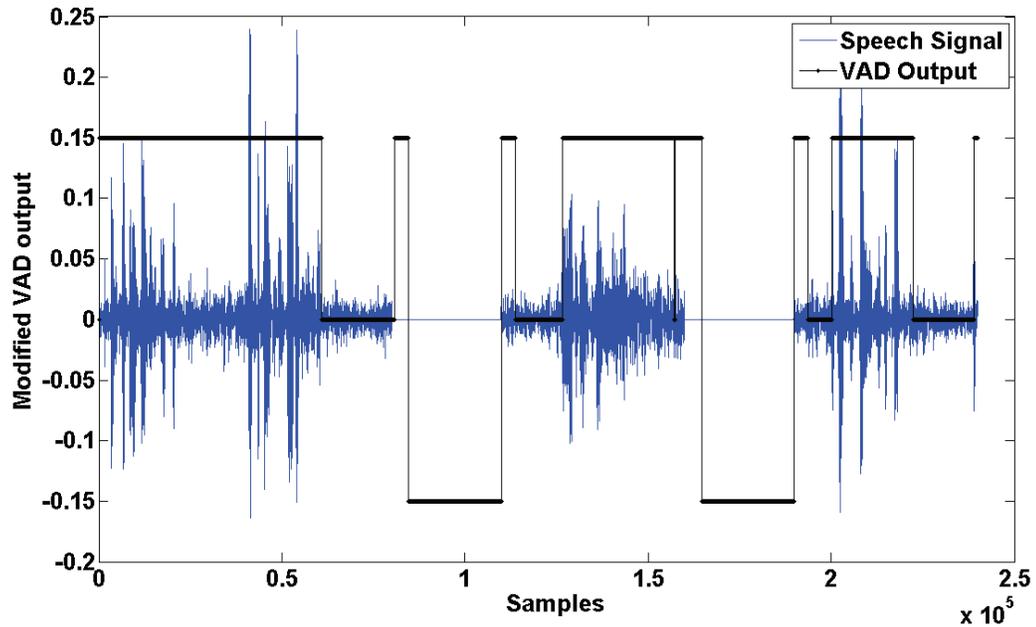

Figure 6.4. Modified Voice Activity Detector (VAD) output on a sample speech signal containing quiet segments (zero indicates noise, positive indicates voice and negative indicates quiet segments).

Figure 6.5 compares the estimated function $\hat{Q}(m)$ defined in (6.19) given by the modified VAD introduced in (5.7) against the actual quiet indicator $Q(m)$ as defined in (6.18). The performance obtained was $P_Q = 0.96$ for the examined speech files which corresponded to a quiet detection accuracy of 96%. The misdetections were caused mainly by the guard time added for having a robust detection as explained earlier. In general, the quiet frames were detected consistently. In other words, the guard time was added in order to increase the classifier's ability to identify negative or non-quiet frames. As shown in Figure 6.5, a specificity (ratio of true negatives to the total number of actual negatives) 100% was achieved at the expense of having about 75% of sensitivity (ratio of true positives to the total number of actual positives).



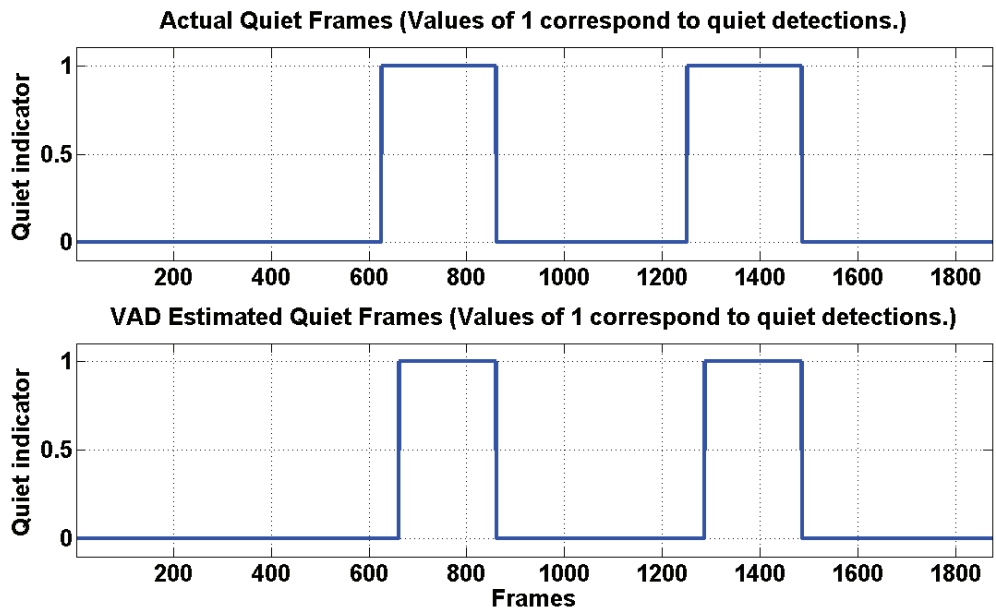

Figure 6.5. Actual and estimated quiet detection.

This was done for quiet detection as it was desired not to turn off suppression in actual non-quiet frames of the input signal. The coefficient $k_Q$ was chosen to maximize the total accuracy defined in (6.20) while still providing 100% specificity. This was consistently generated with a wide range of small $k_Q$ values. In our simulations, the value of $k_Q = 0.01$ was used.

To train the GMM classifier for music/noise classification, four types of noise namely street, car, restaurant and mall were considered to form samples of the noise class and music pieces played by four instruments of piano, guitar, saxophone and violin, were considered to form samples of the music class. These musical instruments were chosen from categories of different types of instruments to cover common types of musical sounds. Each sample in either music or noise class was approximately a one-minute long wave file with a sampling frequency of 22050 Hz. MFCC features were extracted to form the 26-dimensional feature vector for classification. A



total of 75,000 sample frames formed the classification dataset. 80 percent of the data was used for training and the rest of the data was used for testing. Table 6.7 shows the confusion matrix obtained when performing the classification for frames independent of the decision taken for previous frames. As done in (Gopalakrishna, et al. 2012), a majority voting strategy based on the decision taken for previous frames was considered in order to increase the reliability of the classification. In other words, any class change was ignored if it was not been detected in a stable manner which meant taking a majority voting decision of the past 20 classification decisions. After considering the majority voting strategy, 100% classification of the test samples was obtained as noted in Table 6.7 by the numbers inside the parentheses.

In summary, the inclusion of quiet and music detection capabilities is expected to provide an improved hearing experience for CI users when used in conjunction with the previously introduced noise suppression capability. Addition of such a capability was performed with minimal extra computations in order to maintain the computational efficiency of the pipeline. The detection of quiet condition was achieved by introducing a computationally efficient extension of the VAD component as explained in Section 5.3.1. In addition, a quantitative measure was introduced in Section 6.1.3 in order to evaluate the performance of the modified VAD. The detection of music condition was achieved by using a two-class GMM classifier and the same features used for the noise classification while maintaining the computational efficiency of the pipeline. These extensions were shown to provide reliable outcomes as part of the environment-adaptive noise suppression pipeline.



Table 6.7. Normalized confusion matrix percentages for music/noise classification using single frame decisions; the numbers inside the parentheses indicate the classification rates after using the majority voting strategy.

|  |  | **Identified Class** | |
|  |  | *Noise* | *Music* |
| **True Class** | *Noise* | 99% (**100%**) | 1% (**0%**) |
|  | *Music* | 7% (**0%**) | 93% (**100%**) |

**CHAPTER 7**

**CONCLUSION**

This dissertation has introduced a single-processor speech processing pipeline for Cochlear Implants (CIs) primarily to overcome the synchronization challenge in the utilization of bilateral CIs. The following two aspects of the pipeline have been closely studied:

- *Environment adaptability*: It is known that speech understanding of CI users degrades significantly in noisy environments. Most of the noise suppression algorithms currently used in CIs rely on models that consider assumptions about noise statistics. Since different environments such as restaurant, car, office, mall, etc., have different noise characteristics, these models face limitations in a large variety of noise environments that CI users face in their daily lives. The developed solution in this dissertation provides adaptability to optimized environment-specific parameters.

- *Single-processor processing of binaural speech*: It is known that binaural hearing significantly improves speech intelligibility and provides localization cues. Unfortunately, the benefits of using bilateral CIs are currently limited due to the lack of synchronization between left and right stimulation signals. This dissertation has overcome this synchronization problem by developing a single-processor solution to gain the full benefits of using bilateral CIs.





The developed solution has the following two main attributes:

- It is important for the CI speech processing pipeline to be able to automatically detect the noise environment type on-the-fly and tune the corresponding suppression parameters accordingly. The developed framework uses a noise classifier where data captured by two microphones are combined to detect the environment type for the purpose of adjusting the noise suppression component adaptively. This approach was shown to outperform the fixed noise suppression algorithms.

- Actual deployment of the environment-adaptive speech processing pipeline on resource-limited processors of CI devices poses computational and memory efficiency challenges. In this dissertation, the previously developed unilateral pipeline was extended to the bilateral CIs in a computationally efficient manner without using a second processor to avoid the synchronization problem. Enhanced capabilities to detect the background conditions of quiet, music, and noise were also added to the pipeline in a computationally efficient manner.

More specifically, the main contributions of this dissertation as well as a summary of the experimental results obtained are listed below:

- A unified bilateral speech processing pipeline was developed in which binaural stimulations are generated via a single processor.

- The developed solution is environment-adaptive, meaning that it can be optimally tuned automatically without user intervention for different environments with different noise characteristics.



- No need for a second processor naturally resolves the synchronization problem which is viewed as the most challenging problem in bilateral CIs that has limited gaining the full benefits of having binaural stimulations.

- The developed solution provides a sense of directionality to bilateral CI users and also improved speech quality in different real-world noise environments.

- Not only the parameters of the developed speech processing pipeline can be optimally customized to different noise characteristics and be automatically tuned, the utilized data-driven optimization also allows for individualized Head-Related Transfer Functions (HRTFs) to be optimized to model a user's anthropometric measurements. Data collected by users during their daily experience in different noise environments can be used to optimally customize the parameters, making it possible to customize bilateral CIs to a specific user. This paves the way for a highly customizable pipeline to perform optimally in different noise environments and to be easily fitted to an individual user's anthropomorphic measurements.

- A generalized data-driven bilateral speech processing framework was developed to optimize the environment-adaptive pipeline for binaural stimulations using only a single processor. This framework uses the collected data to optimize both suppression and HRTF gain parameters in any given environment and for a specific user.

- Optimization solutions of the data-driven framework were derived for a large variety of speech distortion criteria suitable for different speech quality and intelligibility applications. The solutions reached are general-purpose and include the commonly-used single and dual-channel speech enhancement as special cases. A number of amplitude



and loudness-weighted distortion criteria were examined and specific solutions were obtained for Weighted-Euclidean, Log-Euclidean and Weighted-Cosh cases.

- The experimental results involving optimized suppression and HRTF gain models showed that speech quality improvements were obtained in 78 test cases conducted in 6 commonly-encountered noise environments and 13 different directions on a large set of speech, HRTF, and noise data recorded in real environments using the same BTE (Behind-The-Ear) microphone worn by Nucleus ESPrit implant users.

- The developed pipeline was shown to be 44% faster than the sequential processing of bilateral signals and requires about 50% less memory. The low computational complexity and memory requirement as well as the ability to provide binaural stimulations via only a single processor make the developed solution suitable for deployment on resource-limited processors of CI devices.

- No need for a second processor not only can offer huge decrease in production costs of bilateral CIs, but also eases the challenging task of synchronization, thus making it possible to get the full benefits of using binaural hearing, such as better localization capabilities and speech intelligibility improvements.

In essence, this dissertation has introduced a novel approach to speech processing pipeline of bilateral CIs. Many possible future works on algorithmic improvements of this pipeline can be performed. Some suggested extensions and improvements of different components of the pipeline to better address real-life challenges faced by CI users include: incorporating music processing algorithms utilizing the music detection component, speech intelligibility improvements in reverberation environments, developing optimized solutions for non-



differentiable distortions, studies of different hearing perception models with respect to speech understanding, and extensions involving the use of multiple microphones.

Additionally, there remain a number of system-level extensions and open research studies in clinical evaluations of the developed framework, including subjective evaluations by hearing impaired CI users, clinical studies and tests of patient-specific and user-customized parameters and using them in fitting CI devices with individualized parameters, and studies to see the effects of using different psychoacoustic models of HRTFs in the developed framework on speech understanding. The data-driven aspect is a unique feature of the developed framework for CI applications. In addition to its substantial benefits,—which were discussed throughout the dissertation— the fact that it allows transforming data collected in daily experience of CI users to customize speech processing parameters by obtaining optimized suppression and HRTF parameters, can be the starting point for many interesting future research works, leading to enhanced real-life experiences of CI users.

REFERENCES


Abramowitz, M., and I. A. Stegun. *Handbook of Mathematical Functions.* Dover, New York.: Ninth Dover printing, 1965.

Algazi, V. R., R. O. Duda, D. M. Thompson, and C. Avendano. "The CIPIC HRTF database." *Presented at the IEEE ASSP Workshop on Applications of Signal Processing to Audio and Acoustics.* 2001. 99-102.

Ali, H., A. Lobo, and P. Loizou. "On the design and evaluation of the PDA-based research platform for electric and acoustic stimulation." *Proceedings of IEEE Int. Conf. on Eng. Med. Biol.* San Diego, 2012. 2493–2496.

Ali, H., A. P. Lobo, and P. C. Loizou. "Design and evaluation of a personal digital assistant-based research platform for cochlear implants." *IEEE Trans. Biomed. Eng. 60*, 2013: 3060-3073.

Chen, J., J. Benesty, and Y. Huang. "Time delay estimation in room acoustic environments: an overview." *EURASIP J. Appl. Signal Process. 26*, 2006: 19.

Ching, T. Y.C., E. Van Wanrooy, and H. Dillon. "Binaural-bimodal fitting or bilateral implantation for managing severe to profound deafness: A review." *Trends Amplification 11*, 2007: 161-92.

Cohen, I. "Noise spectrum estimation in adverse environments: Improved minima controlled recursive averaging." *IEEE Trans. Speech Audio Process. 11*, 2003: 466-475.

Ephraim, Y., and D. Malah. "Speech enhancement using a minimum mean-square error short-time spectral amplitude estimator." *IEEE Trans. Acoust. Speech Signal Process. 32*, 1984: 1109-1121.

Ephraim, Y., and D. Malah. "Speech enhancement using a minimum mean-square error-log-spectral amplitude estimator." *IEEE Trans. Acoust. Speech Sign. Proces. 33*, 1985: 443-445.

Erkelens, J., and R. Heusdens. "Tracking of nonstationary noise based on data-driven recursive noise power estimation." *IEEE Trans. Audio, Speech Lang. Process. 16*, 2008: 1112-1123.







Erkelens, J., J. Jensen, and R. Heusdens. "A data-driven approach to optimizing spectral speech enhancement methods for various error criteria." *Speech Commun. 49*, 2007: 530-541.

Fetterman, B., and E. Domico. "Speech recognition in background noise of cochlear implant patients." *Otolaryngol. Head Neck Surg. 126*, 2002: 257-263.

Fingscheidt, T., S. Suhadi, and S. Stan. "Environment-optimized speech enhancement." *IEEE Trans. Audio, Speech, Lang. Process. 16*, 2008: 825–834.

Gopalakrishna, V., N. Kehtarnavaz, and P. Loizou. "A recursive wavelet-based strategy for real-time cochlear implant speech processing on PDA platforms." *IEEE Trans. Biomed. Eng. 57*, 2010a: 2053-2063.

Gopalakrishna, V., N. Kehtarnavaz, and P. Loizou. "Real-time implementation of wavelet-based advanced combination encoder on PDA platforms for cochlear implant studies." *Proceedings of IEEE Int. Conf. on Acoust., Speech, and Sign. Proces.* 2010b. 1670-1673.

Gopalakrishna, V., N. Kehtarnavaz, P. Loizou, and I. Panahi. "Real-time automatic switching between noise suppression algorithms for deployment in cochlear implants." *Proceedings of IEEE Int. Conf. on Eng. Med. Biol.* Buenos Aires, 2010.

Gopalakrishna, V., N. Kehtarnavaz, T. S. Mirzahasanloo, and P. C. Loizou. "Real-time automatic tuning of noise suppression algorithms for cochlear implant applications." *IEEE Trans. Biomed. Eng. 59*, 2012: 1691-1700.

Hansen, J. H.L., and B. L. Pellom. "An effective quality evaluation protocol for speech enhancement algorithms." *Proceedings of Inter.Conf.on Spoken Language Processing 7.* 1998. 2819-2822.

Hu, Y., and P. C. Loizou. "Evaluation of objective quality measures for speech enhancement." *IEEE Trans. Audio Speech Lang. Process. 16*, 2008: 229-238.

Hu, Y., and P. C. Loizou. "Subjective comparison and evaluation of speech enhancement algorithms." *Speech Commun. 49*, 2007: 588-601.

Hu, Y., and P. C. Loizou. "Subjective comparison of speech enhancement algorithms." *Proceedings of IEEE Inter. Conf. on Acoustics, Speech and Signal Process.* 2006.

Hu, Y., P. Loizou, N. Li, and K. Kasturi. "se of a sigmoidal-shaped function for noise attenuation in cochlear implants." *J. Acoust. Soc. Am. 128*, 2007: 128-134.

IEEE Subcommittee. "IEEE recommended practice for speech quality measurements." *IEEE Trans. Audio and Electroacoust. AU-17*, 1969: 225-246.





ITU. "Perceptual evaluation of speech quality (PESQ), and objective method for end-to-end speech quality assessment of narrowband telephone networks and speech codecs." *ITU-T rec. P. 862.* 2000.

Kitawaki, N., H. Nagabuchi, and K. Itoh. "Objective quality evaluation for low-bit-rate speech coding systems." *IEEE J Sel Areas Commun 6*, 1988: 242-248.

Kokkinakis, K., and P. C. Loizou. "Multi-microphone adaptive noise reduction strategies for coordinated stimulation in bilateral cochlear implant devices." *J. Acoust. Soc. Am. 127*, 2010: 3136-3144.

Kühn-Inacker, H., W. Shehata-Dieler, J. Müller, and J. Helms. "ilateral cochlear implants: A way to optimize auditory perception abilities in deaf children?" *Int. J. Pediatr. Otorhinolaryngol. 68*, 2004: 1257-66.

Litovsky, R. Y., et al. "Bilateral cochlear implants in adults and children." *Archives of Otolaryngol. Head Neck Surg. 130*, 2004: 648-55.

Litovsky, R. Y., P. M. Johnstone, and S. P. Godar. "Benefits of bilateral cochlear implants and/or hearing aids in children." *Int. J. Audiol. 45*, 2006: S78-S91.

Loizou, P. C. "Speech processing in vocoder-centric cochlear implants." *Adv. Otorhinolaryngol. 64*, 2006: 109-143.

Loizou, P. C. "Speech quality assessment." *Stud. Comput. Intell. 346*, 2011: 623-654.

Loizou, P. "Speech enhancement based on perceptually motivated Bayesian estimators of the magnitude spectrum." *IEEE Trans. Speech Audio Process. 13*, 2005: 857-869.

Loizou, P. *Speech Enhancement: Theory and Practice.* CRC Press, 2007.

Loizou, P., A. Lobo, and Y. Hu. "Subspace algorithms for noise reduction in cochlear implants." *J. Acoust. Soc. Am. 118*, 2005: 2791-2793.

Lotter, T., and P. Vary. "Speech enhancement by MAP spectral amplitude estimation using a super-gaussian speech model." *EURASIP J. Appl. Signal Process. 2005*, 2005: 1110-1126.

Martin, R. "Noise power spectral density estimation based on optimal smoothing and minimum statistics." *IEEE Trans. Speech Audio Process. 9*, 2001: 504-512.

Mirzahasanloo, T., and N. Kehtarnavaz. "A generalized data-driven speech enhancement framework for bilateral cochlear implants." *Proceedings of IEEE Int. Conf. on Acoust. Speech Signal Process.* 2013a. 7269-7273.





Mirzahasanloo, T. S., and N. Kehtarnavaz. "Real-time dual-microphone noise classification for environment-adaptive pipelines of cochlear implants." *Proceedings of IEEE Int. Conf. on Eng. Med. Biol.* 2013b.

Mirzahasanloo, T., and N. Kehtarnavaz. "A generalized speech enhancement framework for bilateral cochlear implants using a single processor." In *Cochlear Implants: Technological Advances, Psychological/Social Impacts and Long-Term Effectiveness*. Nova Science Publishers, 2014.

Mirzahasanloo, T., N. Kehtarnavaz, and I. Panahi. "Adding quiet and music detection capabilities to FDA-approved cochlear implant research platform." *Proceedings of 8th International Symposium on Image and Signal Processing and Analysis.* 2013.

Mirzahasanloo, T., N. Kehtarnavaz, V. Gopalakrishna, and P. Loizou. "Environment-adaptive speech enhancement for bilateral cochlear implants using a single processor." *Speech Commun, 55*, 2013: 523-534.

Mirzahasanloo, T., V. Gopalakrishna, N. Kehtarnavaz, and P. Loizou. "Adding real-time noise suppression capability to the cochlear implant PDA research platform." *Proceedings of EEE Int. Conf. on Eng. in Med. and Biol.* San Diego, 2012.

Müller, J., F. Schon, and J. Helms. "Speech understanding in quiet and noise in bilateral users of the MED-EL COMBI 40/40+ cochlear implant system." *Ear Hear. 23*, 2002: 198-206.

Nemer, E., R. Goubran, and S. Mahmoud. "Robust voice activity detection using higher-order statistics in the LPC residual domain." *IEEE Trans. Speech Audio Process. 9*, 2001: 217-231.

Quackenbush, S., T. Barnwell, and M. Clements. *Objective Measures of Speech Quality.* 1988.

Ramírez, J., J. C. Segura, C. Benítez, L. García, and A. Rubio. "Statistical voice activity detection using a multiple observation likelihood ratio test." *IEEE Signal Process. Lett. 12*, 2005: 689-692.

Remus, J., and L. Collins. "The effects of noise on speech recognition in cochlear implant subjects: predictions and analysis using acoustic models,." *URASIP J. Appl. Signal Process.: Special issue on DSP in Hearing Aids and Cochlear Implants 18*, 2005: 2979-2990.

Senn, P., M. Kompis, M. Vischer, and R. Haeusler. "Minimum audible angle, just noticeable interaural differences and speech intelligibility with bilateral cochlear implants using clinical speech processors." *Audiol. Neurotol. 10*, 2005: 342–352.





Stadtschnitzer, M., T. Pham, and T. Chien. "Reliable voice activity detection algorithms under adverse environments." *Proceedings of IEEE Second Int. Conf. Commun. Electron.* 2008. 218–223.

Tribolet, J., P. Noll, B. McDermott, and R. E. Crochiere. "A study of complexity and quality of speech waveform coders." *Proceedings of IEEE Int. Conf. Acoust., Speech, Signal Process.* 1978.

van Hoesel, R. J. M. "Sensitivity to binaural timing in bilateral cochlear implant users." *J. Acoust. Soc. Am. 121*, 2007: 2192.

Van Hoesel, R. J.M. "Exploring the benefits of bilateral cochlear implants." *Audiol. Neurootol. 9*, 2004: 234-46.

Van Hoesel, R. J.M., and R. S. Tyler. "Speech perception, localization, and lateralization with bilateral cochlear implants." *J. Acoust. Soc. Am. 113*, 2003: 1617-30.

Zwicker, E. "Subdivision of the audible frequency range into critical bands." *J. Acoust. Soc. Am. 33*, 1961: 248-248.


**VITA**

Taher Shahbazi Mirzahasanloo was born in Meshkin-Shahr, Iran, in 1983. He received his B.Sc. and M.Sc. degrees in Electrical Engineering, from University of Tabriz in 2005, and University of Tehran in 2008. He worked for three years as a researcher at Institute for Research in Fundamental Sciences, School of Cognitive Sciences, Tehran. He joined the Signal and Image Processing Lab at The University of Texas at Dallas for his PhD in Electrical Engineering, in 2010. His research interests are Machine Learning, Artificial Intelligence, Speech and Audio Signal Processing.